\documentclass[preprint]{aastex62}

\hypersetup{linkcolor=red,citecolor=green,filecolor=cyan,urlcolor=magenta}

\graphicspath{{./}{figures/}{figures/cornerPlots/}{figures/stackedDetLimGridPlots/}}

\shorttitle{NICMOS KPI I: Catalogue of BDs}
\shortauthors{Factor \& Kraus}

\begin{document}

\title{NICMOS Kernel-Phase Interferometry I: \\Catalogue of Brown Dwarfs Observed in F110W and F170M}

\correspondingauthor{Samuel M. Factor}
\email{sfactor@utexas.edu}

\author[0000-0002-8332-8516]{Samuel M. Factor}
\affil{Department of Astronomy, The University of Texas at Austin, Austin, TX 78712, USA}

\author[0000-0001-9811-568X]{Adam L. Kraus}
\affil{Department of Astronomy, The University of Texas at Austin, Austin, TX 78712, USA}

\begin{abstract}

Filling out the dearth of detections between direct imaging and radial velocity surveys will test theories of planet formation and (sub)stellar binarity across the full range of semi-major axes, connecting formation of close to wide separation gas giants and substellar companions. Direct detection of close-in companions is notoriously difficult: coronagraphs and point spread function (PSF) subtraction techniques fail near the $\lambda/D$ diffraction limit. 
We present a new faint companion detection pipeline called Argus which analyzes kernel phases, an interferometric observable analogous to closure phases from non-redundant aperture masking but utilizing the full unobscured telescope aperture. 
We demonstrate the pipeline, and the power of interferometry, by performing a companion search on the entire \emph{HST/NICMOS} F110W and F170M image archive of 114 nearby brown dwarfs (observed in 7 different programs). Our pipeline is able to detect companions down to flux ratios of $\sim10^2$ at half the classical diffraction limit. We discover no new companions but recover and refine astrometry of 19 previous imaging companions (two with multiple epochs) and confirm two previous kernel-phase detections. 
We discuss the limitations of this technique with respect to non-detections of previously confirmed or proposed companions. 
We present contrast curves to enable population studies to leverage non-detections and to demonstrate the strength of this technique at separations inaccessible to classical imaging techniques. 
The binary fraction of our sample ($\epsilon_b=14.4^{+4.7}_{-3.0}$\%) is consistent with previous binary surveys, even with sensitivity to much tighter separation companions.

\end{abstract}

\section{Introduction} \label{sec:intro}

The detection of companions to stars---both planets and stellar binaries---has traditionally relied on three methods: radial velocities (RVs), transits/eclipses, and direct imaging. Transits and RVs are both limited in their sensitivity to companions at large semimajor axes; transits are increasingly improbable for companions that are distant from their host star, and RV surveys must extend for at least one orbital period before a detection can be confirmed. In contrast, direct-imaging surveys are more sensitive to objects at larger distances from their host star, and hence they offer a singular view into outer solar systems and the peak of the binary semimajor-axis distribution. However, there is often a gap between these two regimes, inside the inner working angle of direct imaging and outside the regime where transits and RVs can efficiently survey. Filling this gap would offer a crucial new view of both planet formation/evolution and stellar multiplicity.

Binary systems pose stringent tests for models of star and brown dwarf formation, as binaries are a common outcome. A successful formation theory should replicate the observed frequency, semimajor axis distribution, and mass function \citep[e.g.][]{Duchene2013} as well as the detailed orbital parameters \citep{Dupuy2011}, all as a function of system mass. The semimajor axis distribution can provide insight into the size/density evolution of the originating prestellar core, while the companion mass function results from the accretion history. However, it is unclear if the peak of the semimajor axis distribution of brown dwarf binaries has even been resolved  \citep[e.g.][]{Burgasser2006}, particularly among populations of newly forming binaries with known formation environments \citep[e.g.][]{Kraus2012a}. Spectral synthesis might indicate that a significant number of binaries remain unresolved \citep{BardalezGagliuffi2015}, an assertion that could be tested with observations at higher spatial resolution.

While the use of coronagraphs and PSF subtraction techniques can improve detection limits, these techniques still do not achieve sufficient resolution. Imperfections in the optical path (and AO correction for ground based telescopes) introduce ``speckles" which can be misinterpreted as companions. These speckles can be corrected based on temporal or chromatic behavior (i.e. ADI, SDI, and LOCI) at large angular separations, but those methods fail near $\lambda/D$. Interferometric analysis takes advantage of the wave nature of light and can be used to produce a more stable and calibrated PSF, reject speckle noise and detect companions with high contrast at or even below the diffraction limit. Rather than subtracting off the PSF, these techniques uses the information contained in it to infer the geometry of the source. The discovery of the candidate newly forming giant planet LkCa15b by \citet{Kraus2012,Sallum2015} demonstrates the power of such techniques.

The most common interferometric analysis technique for single-aperture telescopes, non-redundant aperture masking interferometry \citep[NRM or AMI;][]{Tuthill2006,Tuthill2012}, places a mask in the pupil plane, transforming a large single aperture into a sparse interferometer. Such a configuration produces an interferogram, or 2-D fringe pattern, which contains information about phase, in addition to intensity. Rather than simply looking for a companion in the image, an observable called a closure-phase is used to compare to a model image. Closure-phases are independent of the path length errors which cause speckles and thus contain true source phase information. NRM has been used for numerous applications in binary demographics and orbital analyses \citep{Ireland2008,Kraus2008,Dupuy2017} and substellar/planetary companion discovery \citep{Lloyd2006,Kraus2012,Sallum2015}. 

A similar observable can be used with a full, rather than a masked, aperture. This technique models the full aperture as a grid of small apertures corresponding to a redundant set of baselines. Using simple linear algebra outlined below, \citet{Martinache2010} derived an operator which, when applied to the sampled phases, produces observables called kernel phases which are again independent of phase errors in each baseline. Diffraction limited observations are required, either from space or using ground based AO systems, as the phase errors \emph{must} be small in order to properly account for the redundancies in the baselines. 

While both methods are powerful tools for super-resolution imaging, NRM faces an observational limit. Since NRM uses a sparse mask, only $\sim5\%$ of the light reaches the detector, imposing a flux limit on possible targets. Kernel-phase interferometry (KPI), on the other hand, utilizes the full telescope aperture. Thus it can achieve similar detection limits in a fraction of the time and can be applied to dimmer sources where NRM is not feasible. As no specialized mask is required, KPI can also be performed on archival data as long as the diffraction pattern is properly sampled. This allows the characterization of a companion's orbit using standard astrometric comparison of archival and current data. 

This technique has been applied only a few times (relative to classical imaging techniques) since the original publication by \citet{Martinache2010}. \citet{Pope2016}, \citet{Kammerer2019}, \citet{Wallace2020}, and \citet{Kammerer2021} all analyzed ground based AO imaging datasets (from Palomar/PHARO, VLT/NACO, Keck/NIRC2 and VLT/VISIR-NEAR, respectively) while \citet{Laugier2019} focused on recovering information from saturated \emph{HST/NICMOS} camera 2 images. Analysis by \citet{Pope2013} of \emph{HST/NICMOS} camera 1 images of brown dwarfs from \citet{Reid2006,Reid2008} proposed five new candidate companions (and four marginal detections) and recovered all previously known companions. Clearly the technique is underutilized and merits further exploration. 

This work analyzes the same two NICMOS data sets as \citet{Pope2013} as well as 5 more sets of archival observations. We introduce a new KPI pipeline named Argus\footnote{The open source python pipeline is available on GitHub: \url{https://github.com/smfactor/Argus}.\label{footnote}} \citep{argus} with a careful treatment of phase errors, calibration, model comparison, and detection limits. We recover 19 observations of 17 previously known companions, including one of the new binaries found by \citet{Pope2013} (as well as one more which we marginally recover). We do not confirm any of their marginal detections. We present refined astrometry and photometry for the detected companions and discuss the resulting binary statistics. All sources which we do not recover can be explained by KPI's insensitivity to wide companions or low SNR imaging. 

This work is laid out as follows. Section \ref{sec:obs} outlines the instrument and archival observations analyzed in this work. Section \ref{sec:meth} details the kernel-phase technique and our new pipeline. Section \ref{sec:res} presents the results of our companion search with discussions of detections and non-detections. In section \ref{sec:disc} we discuss the binary fraction of our sample, and put our KPI analysis in context. In Section \ref{sec:conc} we summarize our conclusions.

\section{Observations} \label{sec:obs}
\subsection{NICMOS Data}\label{sec:nicmos}
In this work we reanalyze archival \emph{HST} observations using the Near Infrared Camera and Multi-Object Spectrometer (NICMOS). NICMOS was installed during servicing mission 2 and operated between 1997 and 1999 (Cycle 7 and 7N) when its solid nitrogen coolant ran out. It was restored to service during servicing mission 3B with a replacement cooling system and operated again from 2002 to 2008 (Cycle 11--16). All data sets analyzed in this work were taken during the second block.

NICMOS operates between 0.8 and 2.5$\mu$m. This work uses the F110W and F170M filters corresponding to a diffraction limited resolution resolution ($\lambda/D$) of 95~mas and 147~mas, respectively. Thus the 43~mas pixels of NIC1 sample the PSF at 2.2 and 3.4 pixels per resolution element, respectively. The detector is $256\times256$ pixels for a field of view of $11\arcsec~\times~11\arcsec$. The X/Y pixel scale ratio for NIC1 is $1.0030$ ($\sim0.1$~mas larger in the X direction). Since we only perform astrometry on the scale of a few pixels, we treat them as square. 

NICMOS contains cold masks which block the thermal emission from the primary mirror edge and telescope spiders. This results in an effective primary mirror diameter of 2.388~m, secondary mirror occultation diameter of 0.8928~m, and a spider arm width of 0.1848~m \citep[dimensions taken from TinyTim;][]{Krist2011}. Due to thermal stress on the NICMOS dewar the cold masks are slightly shifted \citep{Krist1997}. Previous KPI analyses of \emph{HST/NICMOS} data \citep[e.g.][]{Martinache2010,Pope2013} do not implement this shift and we do not as well. Calibration should remove most additional phase noise introduced by the imperfect model, though in the future we could use a gray aperture or characterize the aperture more accurately with a much higher resolution model \citep[e.g.][]{Martinache2020}. 

\subsection{Target Sample}\label{sec:samp}
In this work we analyze the data sets listed in Table~\ref{tab:obs}. These are all the high resolution (NIC1) brown dwarf imaging programs utilizing the F110W and F170M filters. These filters are the two most commonly used filters for imaging of brown dwarfs, roughly corresponding to the \emph{J} and \emph{H} bands. Observing set-ups differ between programs but generally consist of two or more dithered exposures in each filter.

Table~\ref{tab:targets} details the properties of the targets analyzed in this work. The sample covers spectral types from late M to T dwarfs at distances ranging from $\sim$5--35 pc. Programs 9833, 10143, and 10879 were general imaging surveys of nearby brown dwarfs while programs 9843, 10247, and 11136 specifically targeted known binary systems. Program 9704 observed a PSF calibrator (spectral type M4V) in F170M to characterize the cold-mask shift and is included since it is a pristine and high SNR PSF. Program 11136 specifically prioritized wavelength coverage in a large number of filters over deep, high SNR, observations. 

\begin{deluxetable}{clccccll}
\tabletypesize{\scriptsize}
\tablecaption{Observations\label{tab:obs}}
\tablehead{\colhead{Prog. ID} & \colhead{P.I.} & \colhead{Cycle} & \colhead{$N_{obj}$} & \colhead{Filters (in addition to F110W, F170M)} & \colhead{$N_{dithers}$} & \colhead{approx. epoch} & \colhead{Publication}}
\startdata
9704 & \citet{Schultz2002} & 11 & 1 & F180M, F222M & 0 & 12/2002 & \citet{Roye2003} \\
9833 & \citet{Burgasser2003c} & 12 & 22 & F090M & 3--6 & 9/2003--7/2004 & \citet{Burgasser2006} \\
9843 & \citet{Gizis2003} & 12 & 2 & F110M, F145M, F170W & 2 & 9--11/2003 & \citet{Burgasser2011} \\
10143 & \citet{Reid2004} & 13 & 56 &  & 2 & 9/2004--6/2006 & \citet{Reid2006} \\
10247 & \citet{Cruz2004a} & 13 & 1 & F090M, F145M, F160W & 4 & 9/2004 & \citet{Burgasser2006} \\
10879 & \citet{Reid2006a} & 15 & 28 &  & 2 & 7/2006--5/2007 & \citet{Reid2008} \\
11136 & \citet{Liu2007} & 16 & 7 & F090M, F108N, F110M, F113N, F145M, F160W & 3--7 & 5--8/2008 & \citet{Dupuy2010} \\
\enddata
\end{deluxetable}

\startlongtable
\begin{deluxetable}{lhlCCChc}
\tabletypesize{\scriptsize}
\tablecaption{Target information \label{tab:targets} }
\tablehead{\colhead{2MASSJ} & \nocolhead{Other Name} & \colhead{SpT (ref.)} & \colhead{Parallax [mas] (ref.)} & \colhead{J [mag] (ref.)} & \colhead{H [mag] (ref.)} & \nocolhead{Gaia ID} & \colhead{HST program}} 
\startdata
  2MASS J00043484-4044058 &               L  362-29 B &       L5+L5 (\tablenotemark{7}) &     82.1 \pm 0.4 \mathrm{ (\tablenotemark{2})} &   13.109 \pm 0.024 \mathrm{ (\tablenotemark{1})} &   12.055 \pm 0.026 \mathrm{ (\tablenotemark{1})} &  Gaia DR2 4996141155411984128 &    10143 \\
  2MASS J00242463-0158201 &                 V* DY Psc &      M9.5V (\tablenotemark{31}) &   79.97 \pm 0.22 \mathrm{ (\tablenotemark{2})} &   11.992 \pm 0.035 \mathrm{ (\tablenotemark{1})} &   11.084 \pm 0.022 \mathrm{ (\tablenotemark{1})} &  Gaia DR2 2541756977144595712 &    10879 \\
  2MASS J00250365+4759191 &    2MASSI J0025036+475919 &        L4: (\tablenotemark{20}) &     23.7 \pm 1.1 \mathrm{ (\tablenotemark{4})} &     14.84 \pm 0.04 \mathrm{ (\tablenotemark{1})} &   13.667 \pm 0.031 \mathrm{ (\tablenotemark{1})} &                               &    10143 \\
  2MASS J00361617+1821104 &           LSPM J0036+1821 &       L3.5 (\tablenotemark{28}) &  114.42 \pm 0.21 \mathrm{ (\tablenotemark{2})} &   12.466 \pm 0.027 \mathrm{ (\tablenotemark{1})} &   11.588 \pm 0.030 \mathrm{ (\tablenotemark{1})} &  Gaia DR2 2794735086363871360 &    10143 \\
  2MASS J00452143+1634446 &   2MASS J00452143+1634446 &     L2beta (\tablenotemark{24}) &   65.02 \pm 0.23 \mathrm{ (\tablenotemark{2})} &   13.059 \pm 0.022 \mathrm{ (\tablenotemark{1})} &   12.059 \pm 0.035 \mathrm{ (\tablenotemark{1})} &  Gaia DR2 2781513733917711616 &    10143 \\
  2MASS J01075242+0041563 &   2MASS J01075242+0041563 &          L8 (\tablenotemark{3}) &         64 \pm 5 \mathrm{ (\tablenotemark{4})} &     15.82 \pm 0.06 \mathrm{ (\tablenotemark{1})} &     14.51 \pm 0.04 \mathrm{ (\tablenotemark{1})} &                               &    10143 \\
  2MASS J01092170+2949255 &    2MASSW J0109216+294925 &       M9.5 (\tablenotemark{12}) &   62.78 \pm 0.24 \mathrm{ (\tablenotemark{2})} &   12.912 \pm 0.021 \mathrm{ (\tablenotemark{1})} &   12.158 \pm 0.024 \mathrm{ (\tablenotemark{1})} &   Gaia DR2 309977380415042432 &    10879 \\
\enddata
\tablerefs{Reference Key 1: \citet{Cutri2003}, 2: \citet{Gaia2018}, 3: \citet{Faherty2012}, 4: \citet{Schneider2014}, 5: \citet{Skrzypek2016}, 6: \citet{Dupuy2012}, 7: \citet{BardalezGagliuffi2015}, 8: \citet{Schmidt2007}, 9: \citet{Burgasser2002}, 10: \citet{Gagne2015}, 11: \citet{Dupuy2017}, 12: \citet{Reid2008a}, 13: \citet{Faherty2009}, 14: \citet{Marocco2013}, 15: \citet{Kirkpatrick2000}, 16: \citet{Gelino2014}, 17: \citet{Smart2013}, 18: \citet{Burgasser2003}, 19: \citet{Cruz2003}, 20: \citet{Konopacky2010}, 21: \citet{Burgasser2000}, 22: \citet{Aller2016}, 23: \citet{Lodieu2005}, 24: \citet{Liu2006}, 25: \citet{Burgasser2003a}, 26: \citet{Leggett2001}, 27: \citet{Tinney2014}, 28: \citet{Reid2000}, 29: \citet{Perryman1997}, 30: \citet{Kirkpatrick1991}, 31: \citet{Faherty2016}, 32: \citet{Reid2006}, 33: \citet{Burgasser2005}, 34: \citet{Chiu2006}, 35: \citet{Burgasser2006a}, 36: \citet{Cardoso2015}, 37: \citet{West2008}, 38: \citet{Metchev2008}, 39: \citet{Reyle2010}, 40: \citet{Burgasser2010}, 41: \citet{Deshpande2012}, 42: \citet{Kirkpatrick2016}, 43: \citet{Liu2016}, 44: \citet{Marocco2015}, 45: \citet{Kirkpatrick2019}, 46: \citet{Best2020}, 47: \citet{Smart2018}, 48: \citet{Manjavacas2013}}
\end{deluxetable}

While previous studies of these data sets have found tight companions, we analyzed all images with no prior knowledge other than a visual inspection for obvious secondary sources. KPI produces improved astrometric precision and searches for tighter and fainter companions around targets previously thought to be single. While \citet{Pope2013} previously analyzed the \citet{Reid2006,Reid2008} observations using KPI, this work uses a new code-base with a detailed treatment of phase errors, calibration, model comparison, and detection limits in addition to analyzing data from 5 more \emph{HST} programs. There are also a few more subtle differences in the analysis which are discussed further below.

\section{Methods} \label{sec:meth}

\subsection{Kernel-phase Analysis}\label{sec:kerph}
The kernel phase derivation was first presented by \citet{Martinache2010} and is summarized here for clarity since the technique is still relatively obscure. KPI works by modeling the full telescope aperture as a grid of sub-apertures. From this grid, a set of baselines (vectors in $uv$ space) can be generated by pairing apertures. Figure~\ref{fig:mask} shows our model of the \emph{HST/NICMOS} aperture and the baselines it samples. Phases (or more generally, complex visibilities) are then sampled from the Fourier transform of the image at each of these points in $uv$ space. Kernel phases can then be constructed based on the geometry of the model aperture using linear algebra outlined below. 

Each measured phase ($\Phi$) is made up of the true source phase ($\Phi_0$) and a combination of phase errors ($\phi$) introduced by pupil path length differences from each of the apertures contributing to the baseline:
\begin{equation}\label{eq:ind-phase}
\Phi=\mathrm{Arg}\left[\sum_{(j,k)} e^{i(\Phi_0+\phi_j-\phi_k)}\right],
\end{equation}
where the pair $(j,k)$ denotes the baselines contributing to the specific visibility. The complex interferometric visibility is thus a superposition of vectors (i.e. the \emph{phases} do not strictly add). If we assume these error terms are small, we can approximate each complex phasor as a linear sum, $e^{i(\phi_j-\phi_k)}\approx1+i(\phi_j-\phi_k)$, allowing us to write the measured phase as a linear combination:
\begin{equation}\label{eq:lin-ind-phase}
\Phi\approx\Phi_0+\frac{1}{r}\sum_{(j,k)} (\phi_j-\phi_k),
\end{equation}
where $r$ is the number of $(j,k)$ pairs for that specific phase measurement (i.e. number of redundant baselines contributing to the visibility). If we stack these equations for each sampled point in $uv$ space, we can write the vector of all measured phase information as
\begin{equation}\label{eq:mat-phase}
\mathbf{\Phi}=\mathbf{\Phi}_0 + \mathbf{R}^{-1}\cdot\mathbf{A}\cdot\mathbf{\phi},
\end{equation}
where the matrix $\mathbf{R}$ is a diagonal matrix with $1/r$ for each visibility, encoding the redundancies in the baselines. The matrix $\mathbf{A}$ is a transfer matrix assembled from 1's and --1's in each row encoding the apertures contributing to each (redundant) baseline (any apertures not involved are 0 and apertures contributing multiple times to a single baseline are summed). For $\phi$ of length $n$ (the number of sub-apertures), and $\Phi$ of length $m$ (the number of unique visibilities), the shape of $\mathbf{R}$ is $m\times m$ and $\mathbf{A}$ is $m\times n$.

Since our goal is to derive a quantity which does not involve the phase error term, we then construct the kernel, $\mathbf{K}$, of $\mathbf{A}$ such that $\mathbf{K}\cdot\mathbf{A}=\mathbf{0}$.  The matrix $\mathbf{K}$ can be calculated using singular value decomposition \citep[SVD,][]{Press2002}. This algorithm gives 
\begin{equation}\label{eq:svd}
\mathbf{A}=\mathbf{U}\cdot\mathbf{W}\cdot\mathbf{V}^T,
\end{equation}
where the matrix $\mathbf{W}$ is a diagonal matrix of positive or zero ``singular values." The rows of $\mathbf{K}$ are then the \emph{columns} of $\mathbf{U}$ corresponding to zero elements of $\mathbf{W}$. After multiplying equation \ref{eq:mat-phase} by $\mathbf{R}$ and then $\mathbf{K}$ we have 
\begin{equation}\label{eq:ker}
\mathbf{K}\cdot\mathbf{R}\cdot\mathbf{\Phi}=\mathbf{K}\cdot\mathbf{R}\cdot\mathbf{\Phi}_0.
\end{equation}
This results in a set of phase-like observables ($\mathbf{K}\cdot\mathbf{R}\cdot\mathbf{\Phi}$), called kernel phases, which are free of instrumental phase errors.

\subsection{Kernel-Phase Pipeline}\label{sec:pipe}

The pipeline can be divided into four steps: aperture characterization, kernel-phase calculation, calibration, and model fitting. During aperture characterization we create a model interferometer that approximates the full aperture of the telescope, measure the associated baselines, and construct the transfer matrices. We then ingest the images and calculate the kernel phases. These kernel phases must then be calibrated to remove systematic errors. Once calibrated, we can then fit a single and binary model to the data and compare the results.

\subsubsection{Aperture Characterization}\label{sec:aper}
\begin{figure*}
\resizebox{\textwidth}{!}{
\includegraphics[height=3cm]{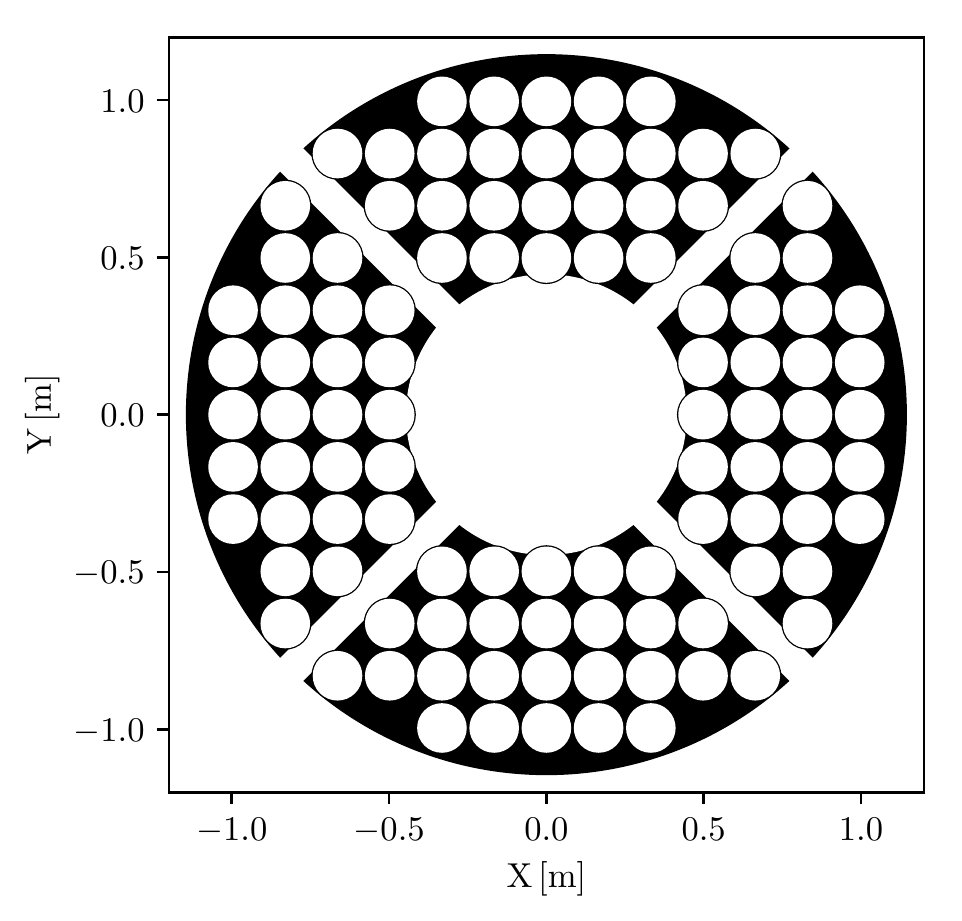}
\includegraphics[height=3cm]{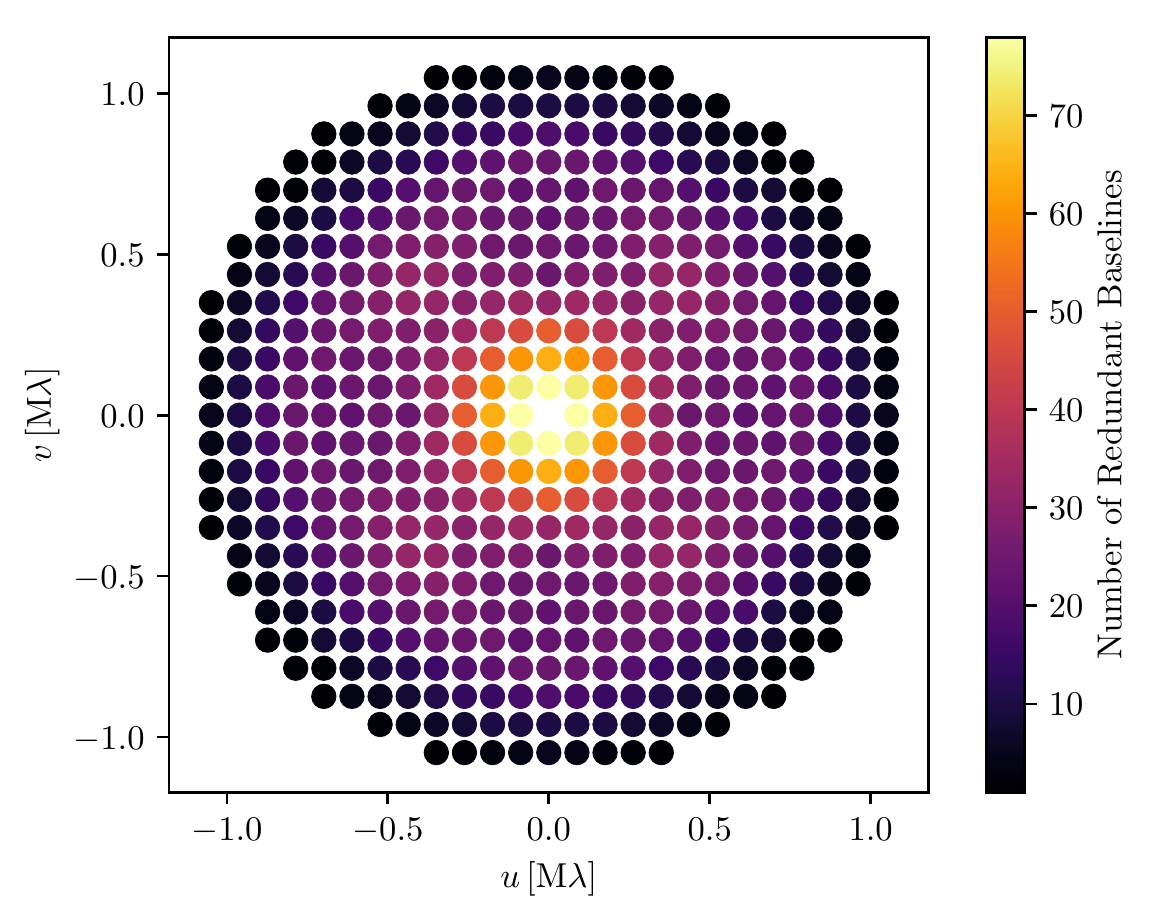}}
\caption{\emph{Left:} HST aperture in black with simulated subapertures on top in white. \emph{Right:} The corresponding baselines (at $1.7\,\mathrm{\mu m}$), color coded by the number of distinct pairs of sub-apertures which contribute to the point. The 104 sub-apertures sample 258 spatial frequencies and generate 206 kernel phases.}\label{fig:mask}
\end{figure*}

First we must construct a set of subapertures that models the full aperture of the telescope. Dimensions of the primary mirror and its obstructions were taken from the TinyTim software package \citep{Krist2011} and are outlined in section \ref{sec:nicmos}. The model used in this work, shown in the left panel of Figure~\ref{fig:mask}, was chosen by balancing a few characteristics: the fraction of the total aperture covered by subapertures ($F_\mathrm{cov}$), the length of the longest baseline ($D_\mathrm{max}$), the phase coverage or number of kernel phases relative to number of phases sampled ($N_\mathrm{ker}/N_\mathrm{B}$), and the number of long baselines relative to the total number of baselines ($N_\mathrm{long}/N_\mathrm{B}$). In general, $F_\mathrm{cov}$ and $D_\mathrm{max}$ trend larger with smaller subaperture size while $N_\mathrm{ker}/N_\mathrm{B}$ and $N_\mathrm{long}/N_\mathrm{B}$ trend smaller. The total number of kernel phases must also be kept to a reasonable number to keep fitting fast and such that we do not try to extract more information than is contained in the image. The theoretical minimum spacing of apertures is set by the size of the image, or in our case the super-Gaussian windowing function (see Section~\ref{sec:kerphCalc}). The field of view, and therefore the smallest spacial frequency sampled by the image, is $25\lambda/D$ and thus the minimum aperture spacing must be at least $D/25$ or 0.096~m. Our chosen spacing is $D$/14.4 or 0.166~m.

From the model of the aperture, the pipeline measures the baselines (right panel of Figure~\ref{fig:mask}), constructs the $\mathbf{A}$ matrix, and then calculates the kernel-phase transfer matrix $\mathbf{K}$. This step is only based on the telescope and sub-aperture geometry, and thus does not need to be repeated for each target. 

\subsubsection{Kernel-Phase calculation}\label{sec:kerphCalc}
The second step is to calculate kernel phases for each target. The pipeline reads in the images and interpolates over bad pixels. Bad pixels are identified from the data-quality flags associated with the individual images. Bad pixels are then iteratively replaced with the median of the 8 neighboring pixels. Other methods of dealing with bad pixels which avoid interpolation may be incorporated in future releases and are discussed further in Section \ref{sec:conc}. The pipeline then finds the flux centroid of the target $(x_\mathrm{cen},y_\mathrm{cen})$ and windows the data with a super-Gaussian ($e^{-(r/\sigma)^4}$) of width $\sigma=25\lambda/D$ ($\sim2.4$ and $\sim3.7$ arc sec in F110W and F170M, respectively) to limit sensitivity to readout noise \citep[as done in][]{Kraus2008}. The images are then Fourier transformed using the basic \texttt{numpy} FFT routines. We then divided the complex visibilities by the position offset of the centroid: $e^{i(x_\mathrm{cen}*u+y_\mathrm{cen}*v)}$. Shifting in the Fourier domain avoids any pixel interpolation which may distort the phase information encoded in the image. 

The phase at each point defined by the aperture model is then measured, along with the local dispersion. The errors associated with kernel phases are derived from the dispersion of the phases measured from the Fourier transform of the image. First, we define the scale on which the $uv$ plane is sampled: $\Delta f=d/2\lambda$. This spatial frequency scale $\Delta f$ is set by the grid spacing of the aperture model $d$  (the diameter of the simulated sub-apertures), and the wavelength $\lambda$. This defines a circle of radius $\Delta f$ around each point in the $uv$ plane corresponding to all possible vectors from any point within the first sub aperture to any point within the second. The phase and associated error are then the mean and standard deviation of all points inside this circle given by the fast Fourier transform. These phases and associated errors are then propagated through the kernel-phase transfer matrix generated in the previous step to obtain the kernel phases and associated errors. 

Sampling a number of phases around the baseline (rather than at a single point) appropriately captures the local dispersion of each phase, but not any systematic phase errors. Many previous applications of this analysis technique have introduced a global systematic error term, added in quadrature to the individual errors, to account for this and bring reduced $\chi^2$ values of the best fit models closer to one. We fit this error term \citep[iteratively, using a basic \texttt{python} fitting routine, e.g. \texttt{scipy.optimize.curve\_fit}][]{SciPy} for each image since, for constant aberration, phase errors scale with $\lambda$ and may vary over the detector.

\begin{figure}
\includegraphics[width=\textwidth]{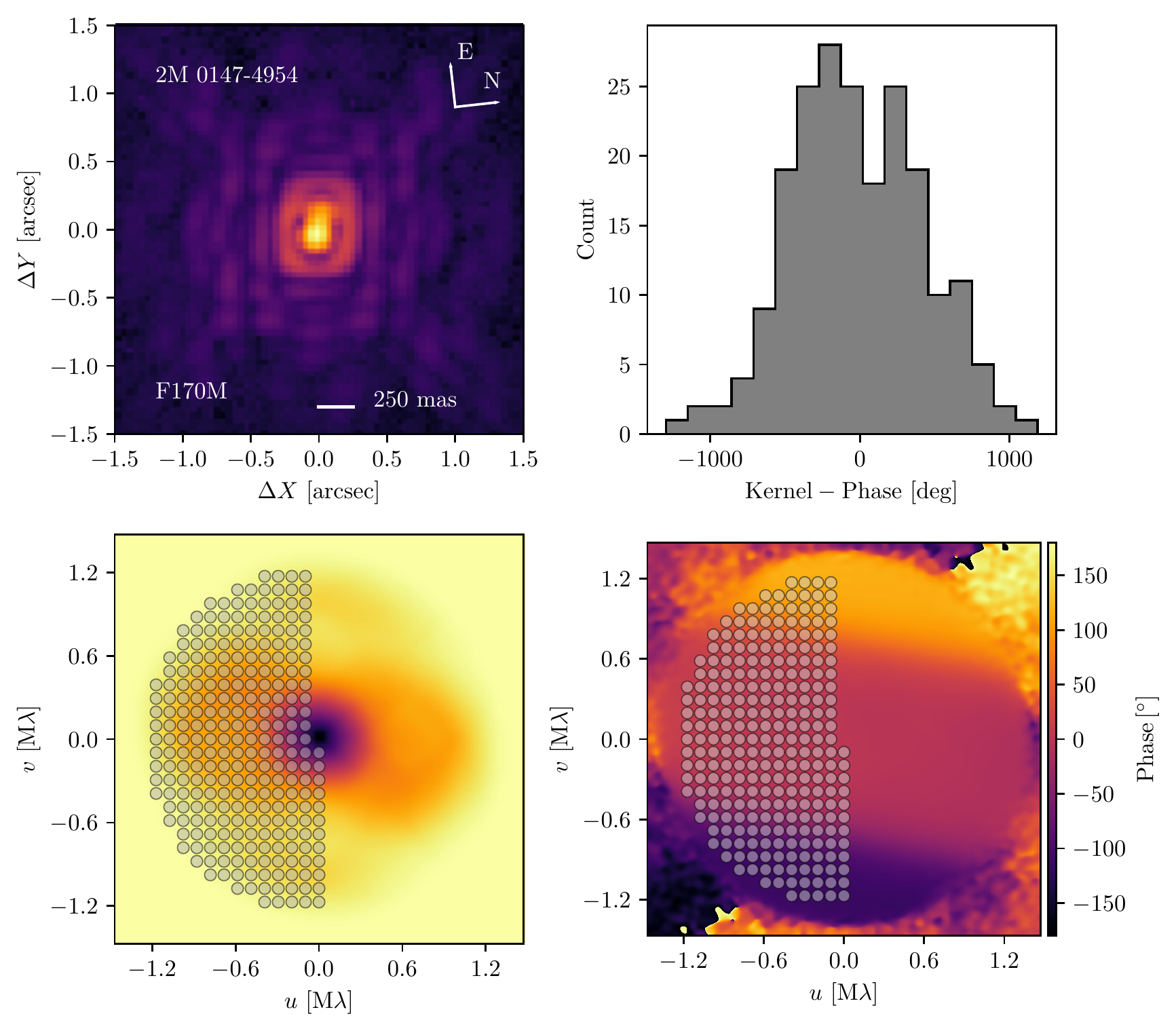}
\caption{The progression from image to kernel phase for an observation of 2MASS J0147-4954, a brown dwarf with a companion at $\sim140$ mas ($\sim 1 \lambda$/D) and ~2:1 contrast in F170M. \emph{Top-left}: \emph{HST/NICMOS} NIC1 image (fourth root scaling). \emph{Bottom row}: Fourier-amplitude (left), and Fourier-phase (right). Gray circles show the spatial frequencies sampled by the model aperture shown in Figure~\ref{fig:mask}. \emph{Top-right} Histogram of measured kernel phases. A point source would have kernel phases of $0^\circ$ (with some noise) which is clearly not the case.} 
\label{fig:imsBin}
\end{figure}

\begin{figure}
\includegraphics[width=\textwidth]{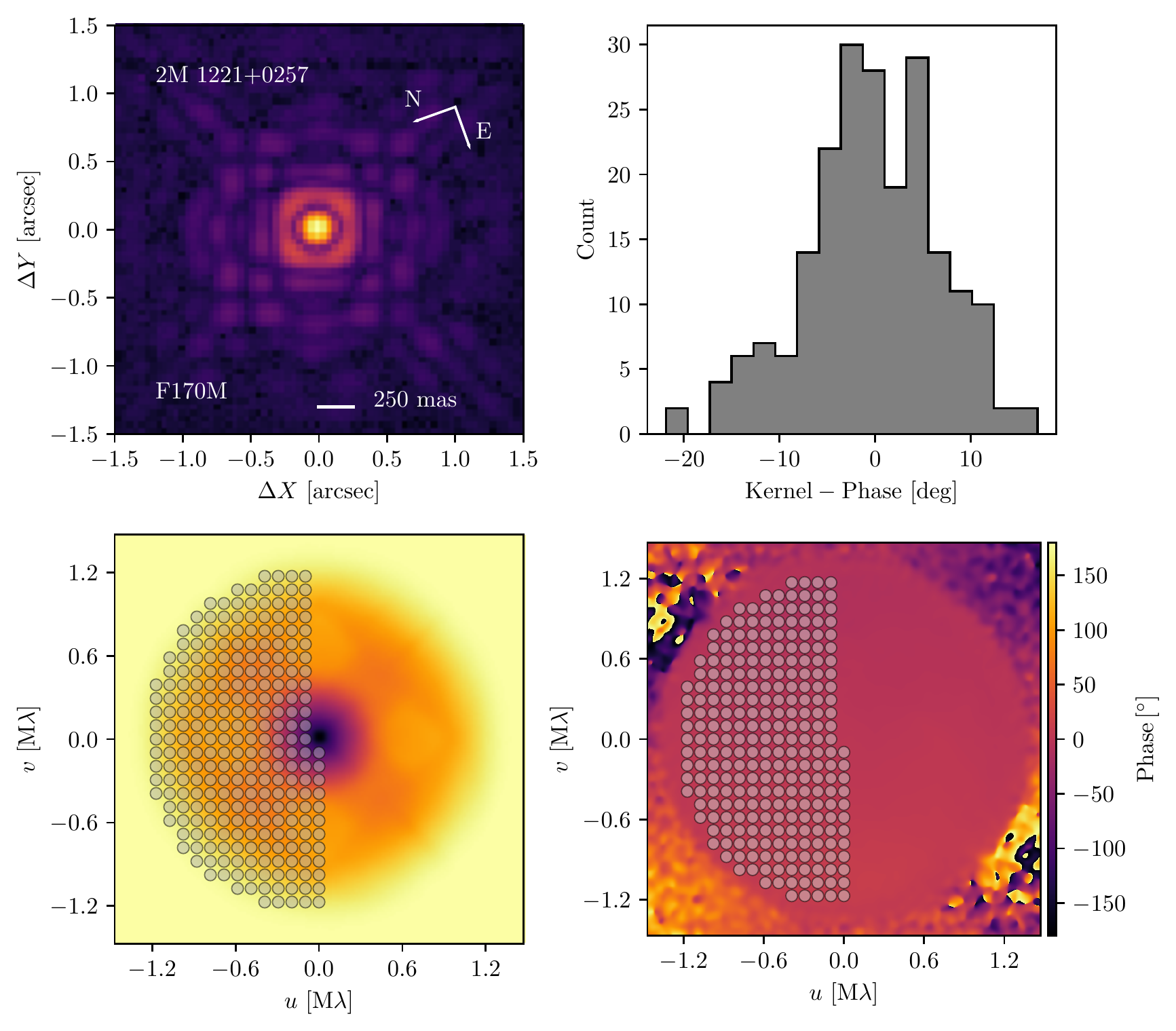}
\caption{Similar to Figure~\ref{fig:imsBin} but for 2MASS J1221+0257, a brown dwarf with no companion. \emph{Top-left}: \emph{HST/NICMOS} NIC1 image (fourth root scaling). \emph{Bottom row}: Fourier-amplitude (left), and Fourier-phase (right). Gray circles show the spatial frequencies sampled by the model aperture shown in Figure~\ref{fig:mask}. \emph{Top-right} Histogram of measured kernel phases. A point source would have kernel phases of $0^\circ$ (with some noise) as is the case with this target.} 
\label{fig:imsSin}
\end{figure}

Figure~\ref{fig:imsBin} shows intermediate steps for 2MASS J0147-4954 at this stage of the analysis: the bad pixel corrected and windowed image, the Fourier transform of the image (amplitude and phase) along with the modeled baselines, and the measured kernel phases. This target is marginally resolved at this wavelength, and a step function, characteristic of a binary, can clearly be seen in the Fourier-phase. The large dispersion of kernel phases in the upper right panel further indicates the presence of a companion. Figure~\ref{fig:imsSin} show the same plots but for 2MASS J1221+0257, a target with no companion. The flat phase and $\sim100\times$ smaller kernel phases indicate that this target does not have a companion. 

\subsubsection{Calibration}\label{sec:cal}
The third step is calibration. The phase signal of a single centered point source should be zero everywhere, producing all zero kernel phases. However, path-length offsets in the telescope and instrument produce non-zero phases and kernel phases (as seen in Figure~\ref{fig:imsSin}). If these offsets are stable, which they are for similar observing setups (and for inherently stable space based observatories), they can be calibrated out. Subtracting off the (non-zero) kernel phases of an observed point source sets the zero-point in the science target such that a single source would now have all zero kernel phases.  

Since no dedicated PSF calibrators were observed in these archival datasets, we use the science targets themselves as calibrators. To create kernel phases which can be used to calibrate science targets, a preliminary fit is run on each target to reject binaries and remove any small centroid offset error. As part of the data ``onboarding" procedure, any images with visual signs of a companion are noted and rejected at this step. The best fit single point source model (x,y position offset) is then subtracted from the kernel phases of each point source to create calibrators. These are then used to calibrate science targets before fitting for binary parameters. Each science target is paired with 5 calibrators, balancing the increased confidence of a detection in multiple calibrators with the increased computation time of fitting multiple times. The results from these 5 independent fits are then used to characterize if a companion is present according to the metric discussed in Section~\ref{sec:detLim}. 

We experimented with multiple calibrators chosen from the most singular sources (those which favored the single point source model over the double model the most), the closest on the detector, and a combination of the two metrics. We chose to use the calibrators located closest to the science target on the detector, as the detection limits (see Section \ref{sec:detLim}) using those calibrators were significantly more sensitive. Targets which are co-located on the detector calibrate each other well since they accumulate similar phase offsets from passing through or reflecting off of the same region of optical elements within the telescope and instrument.

\subsubsection{Model Fitting}\label{sec:fitting}
The fourth and final step is to pass the calibrated kernel phases to a Bayesian inference algorithm which fits a single and a double point source model to the data. The free parameters of the model are a small position offset (to allow for sub-pixel refinement of the centroid) and binary separation, position angle, and contrast ratio. We have chosen to use a Gaussian prior on the centroid (centered at 0 with a standard deviation of 10 mas), $\log$-uniform priors on separation and contrast, and a uniform prior on position angle (the code we use allows a ``wrapped" prior which we turn on for position angle). The $\log$-uniform prior is commonly used as an uninformative prior on scaling parameters like separation and contrast as it allows for a wide range of values with equal probability over each power of 10.

The fitting routine used in this work, PyMultiNest \citep{Buchner2014} uses a nested sampling algorithm which samples the entire prior volume, probabilistically constricting that volume down to the best fit(s) \citep{Feroz2008,Feroz2009,Feroz2013}. Since it samples the entire (allowed) parameter space, it also allows the calculation of the Bayesian evidence which can then be used to calculate Bayes factors, comparing the single and double point source models (see Section \ref{sec:bayes}). 

\begin{figure}
\includegraphics[width=\textwidth]{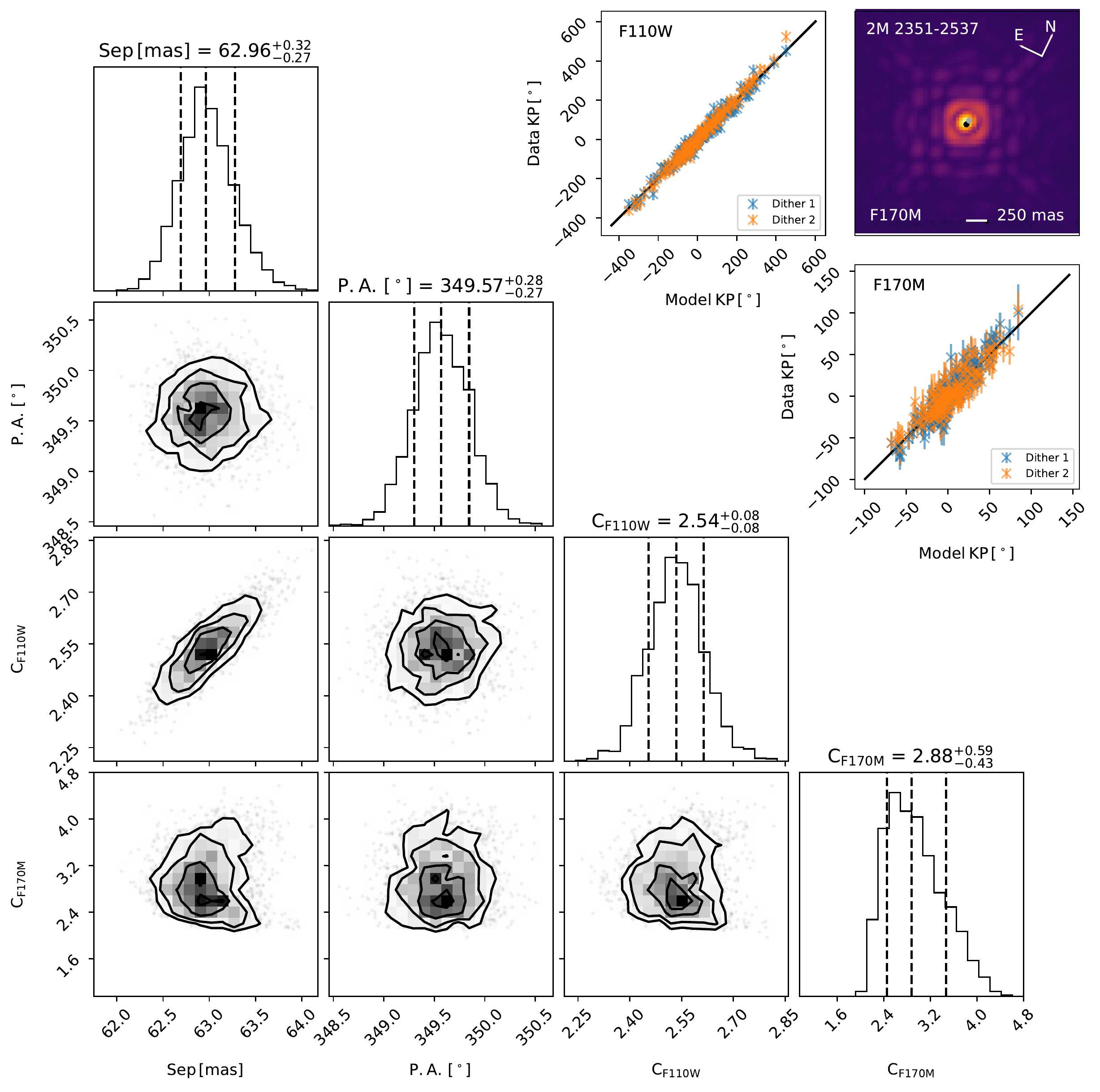}
\caption{Results of fitting a double point source model to observations of 2MASS 2351-2537 (example image shown in the upper right corner with dots showing the position and contrast ratio of the two sources). \emph{Lower Left}: Corner plot showing the posteriors of the four-parameter fit. Dashed lines indicate the median and $\pm 1\sigma$ values (16th, 50th, and 86th percentiles). Top Right: Data kernel phases plotted against the best-fit model kernel phases indicating a good fit. Detection limits for this fit show it is significant at the $>5\sigma$ level and is consistent over multiple calibrators, while the Bayes-factors also indicate a preference for the binary model. Similar figures for each of our detections are shown in Figure Set 4 while marginal and spurious detections are shown and discussed in Sections~\ref{sec:marginal} and \ref{sec:spurious}, respectively. The complete figure set (19 images) is available online at \citet{figSets}. }\label{fig:model}
\end{figure}

Figure~\ref{fig:model} shows the 1D and 2D posterior distributions along with the calibrated kernel phases generated from the data plotted against the best fit binary model for an binary system with a separation well below the diffraction limit, 2MASS J2351-2537. This test case, where the image does not clearly show a companion but the kernel phases plainly indicate the presence of a binary, demonstrates the power of this interferometric analysis, detecting an unresolved binary with a contrast ratio of $\sim2.5:1$ at $\sim0.3\lambda/D$. Similar figures for all of our detected companions are shown in Figure Set 4.

\subsection{Detection Limits}\label{sec:detLim}
Detection limits were calculated in a similar manner to applications of NRM \citep{Kraus2008,Kraus2011}. For each source-calibrator pair, the best fit single point-source model is subtracted from the calibrated kernel phases leaving behind noise (assuming no companion is present). The indices of these kernel phases are then scrambled, creating a new realization of the noise with the same properties. A single point source is then added back in to the scrambled kernel phases at the position of the single point source originally subtracted to replicate the signal of the centroid offset.

We then fit for contrast on a 100x100 grid in separation--PA space. Posteriors from these fits are then combined, marginalizing over PA. Since the indices have been scrambled, these fits are to kernel phases containing no real companions and thus the posteriors indicate what regions of parameter space produce spurious detections due to noise. 

\begin{figure}
\centering
\includegraphics[width=\textwidth]{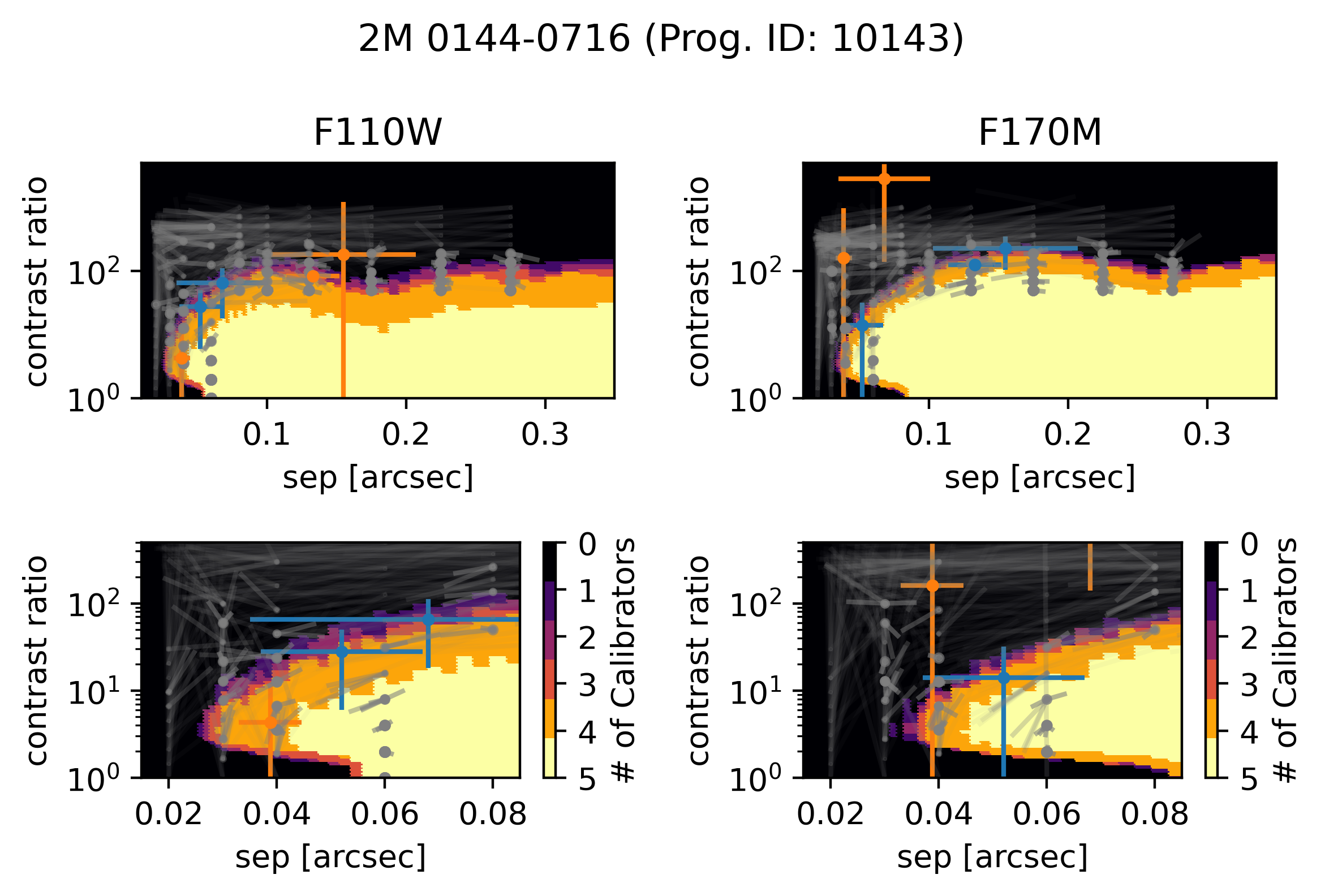}
\caption{Example detection limits on a source with no companion detected. The color scale is the number of calibrators which would significantly ($>5\sigma$) detect a companion as a function of separation and contrast ratio in the two filters (\emph{left} and \emph{right} columns). The \emph{top row} shows the full range of separation, while the \emph{bottom row} zooms in on the extremely close separation regime where there is a decrease in sensitivity to equal brightness companions (due to a decrease in asymmetry). Best fit companion parameters are plotted (with error bars) for the 5 calibrators in blue, if significantly detected, and in orange, if not significantly detected. In this case $<4$ calibrators produce a significant detection and the parameters (including PA) are not consistent with each other so no companion is detected. Overlaid in gray are the results of an injection-recovery test. The injected source is indicated by a circle and is connected to the recovered parameters by a line. Symbol size and opacity is scaled by the proximity of the injected and recovered parameters. Similar plots for the other single sources are shown in Figure Set 5. The complete figure set (83 images) is available online at \citet{figSets}.}
\label{fig:nonSigDetLim}
\end{figure}

\begin{figure}
\centering
\includegraphics[width=\textwidth]{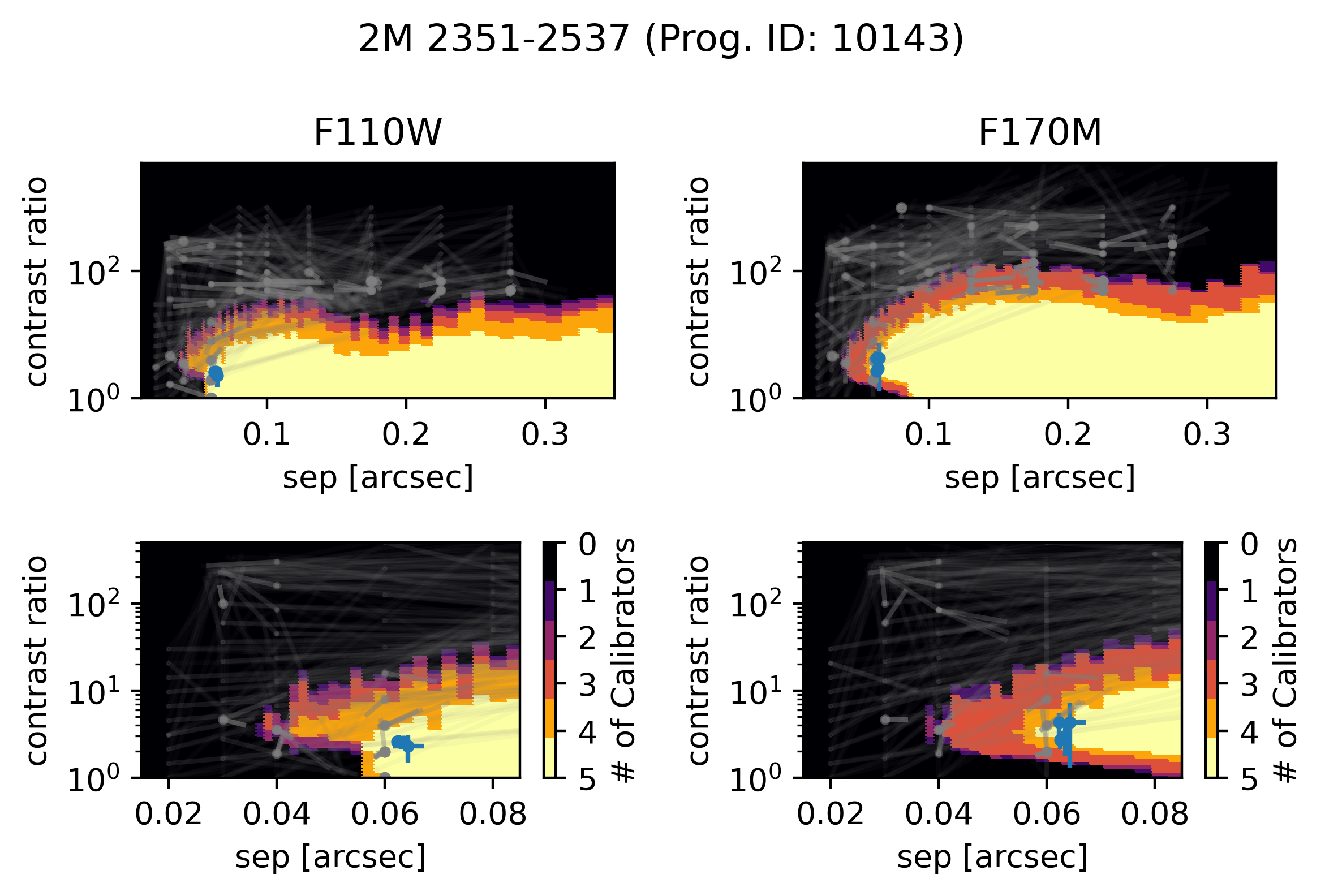}
\caption{Similar to Figure~\ref{fig:nonSigDetLim} but for a significantly detected source. In this case all calibrators produce significant detections and the best fit parameters are consistent with each other.  Similar plots for significant detections are shown in Figure Set 6. The complete figure set (19 images) is available online at \citet{figSets}.}
\label{fig:sigDetLim}
\end{figure}

\begin{figure}
\centering
\includegraphics[width=\textwidth]{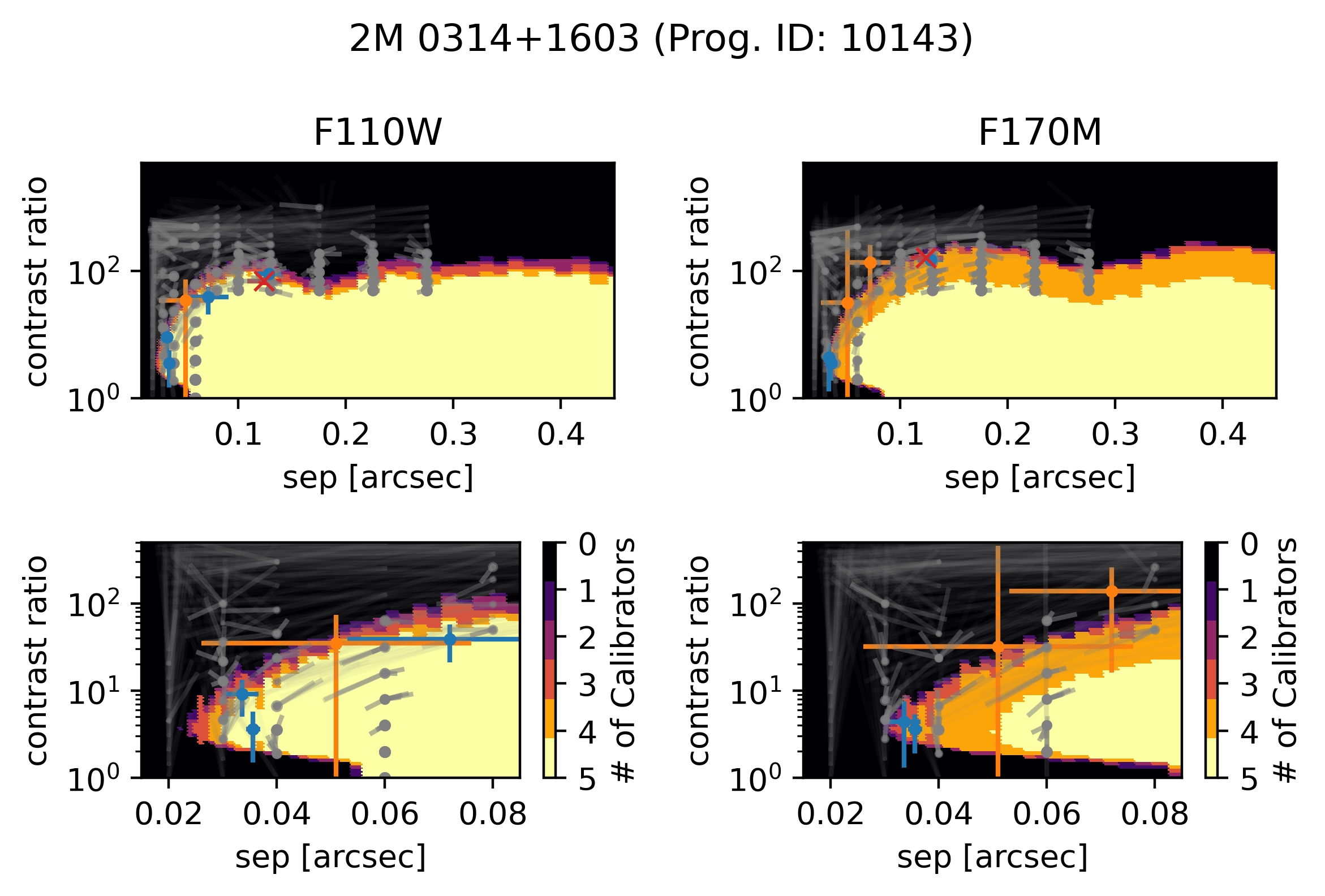}
\caption{Similar to Figure~\ref{fig:nonSigDetLim} but for a notable non-detection. In this case 2M 0314+1603 was marginally detected by \citet{Pope2013}. Their best fit parameters are indicated by the red X. One of our calibrators roughly recovers their values but they are not consistent between enough calibrators to consider this detection significant. Similar plots for notable non-detections are shown in Figure Set 7. The complete figure set (11 images) is available online at \citet{figSets}.}
\label{fig:noteNonDetLim}
\end{figure}

The confidence levels are then drawn such that the $X$\% confidence contour contains $100-X$\% of the posterior moving from least dense to most dense areas. This is done using a Kernel Density Estimate of the posterior as a function of separation and contrast. Example detection limits are shown in Figures~\ref{fig:nonSigDetLim}, \ref{fig:sigDetLim}, and \ref{fig:noteNonDetLim} for a target with no companion, a confident detection, and a notable non-detection, respectively. Corresponding figures for the rest of the sources are shown in Figure Sets 5, 6, and 7, respectively with Marginal and Spurious detections shown in Figure Set 9.

We also run an injection/recovery grid to verify the NRM style scramble-and-fit detection limits which is overlaid in those figures. This mainly serves as a check to confirm that the scramble-and-fit detection limits are correctly positioned with respect to the companions that the pipeline is able to recover.

A fit companion is considered a confident detection if it is detected at $>5\sigma$ confidence in at least four of the five calibrators and the binary companion parameters are consistent between every pair of calibrators (with significant detections) in position and contrast within 5 and 1 times the $1\sigma$ error bars in the fits, respectively. These limits were selected in order to maximize the number of known companions we recover while minimizing the number of spurious detections. Decreasing the number of calibrators which a companion must be significantly detected in adds spurious detections, all of which are detections in only a single filter. Increasing the precision with which fits using different calibrators must be consistent with each other eliminates detections which we know are real from the literature. The fact that the positional error bars must be inflated while the contrast error bars do not suggests that the calibration uncertainty is much larger with respect to the statistical uncertainty in position than in contrast. \citet{Ceau2019} also discuss a hypothesis testing framework which we discuss further in Section~\ref{sec:scrambleVsBF}.

\subsection{Bayesian Model Comparison}\label{sec:bayes}

Since we are using PyMultiNest to fit our data, we also calculate the Bayesian evidence for the single and double point source models. This value can be used to compare the two models in a purely Bayesian manner, irrelevant of the best fit parameter values. Bayes's theorem (including the hypothesis along with the parameters) states 
\begin{equation}\label{eq:bayes}
P(\theta|D,H)=\frac{P(D|\theta,H)P(\theta|H)}{P(D|H)},
\end{equation}
where $\theta$ is the model parameters, $D$ is the data, $H$ is the hypothesis (or specific model), $P(\theta|D,H)$ is the posterior, $P(D|\theta,H)$ is the likelihood, $P(\theta|H)$ is the prior, and $P(D|H)$ is the Bayesian evidence. The Bayesian evidence can be thought of as a normalization factor for the numerator, since the posterior distribution must integrate to 1. The Bayesian evidence is ignored by typical MCMC routines, since the integral (over all parameter values) is difficult to perform, and thus the equality is reduced to a proportionality. Since nested sampling algorithms \citep[e.g.][]{Buchner2014} sample the entire prior volume, they can calculate the Bayesian evidence. The ratio of the Bayesian evidence for two different models, called the Bayes factor, can then be used to compare two different models. 

\begin{figure}
    \centering
    \includegraphics[width=\textwidth]{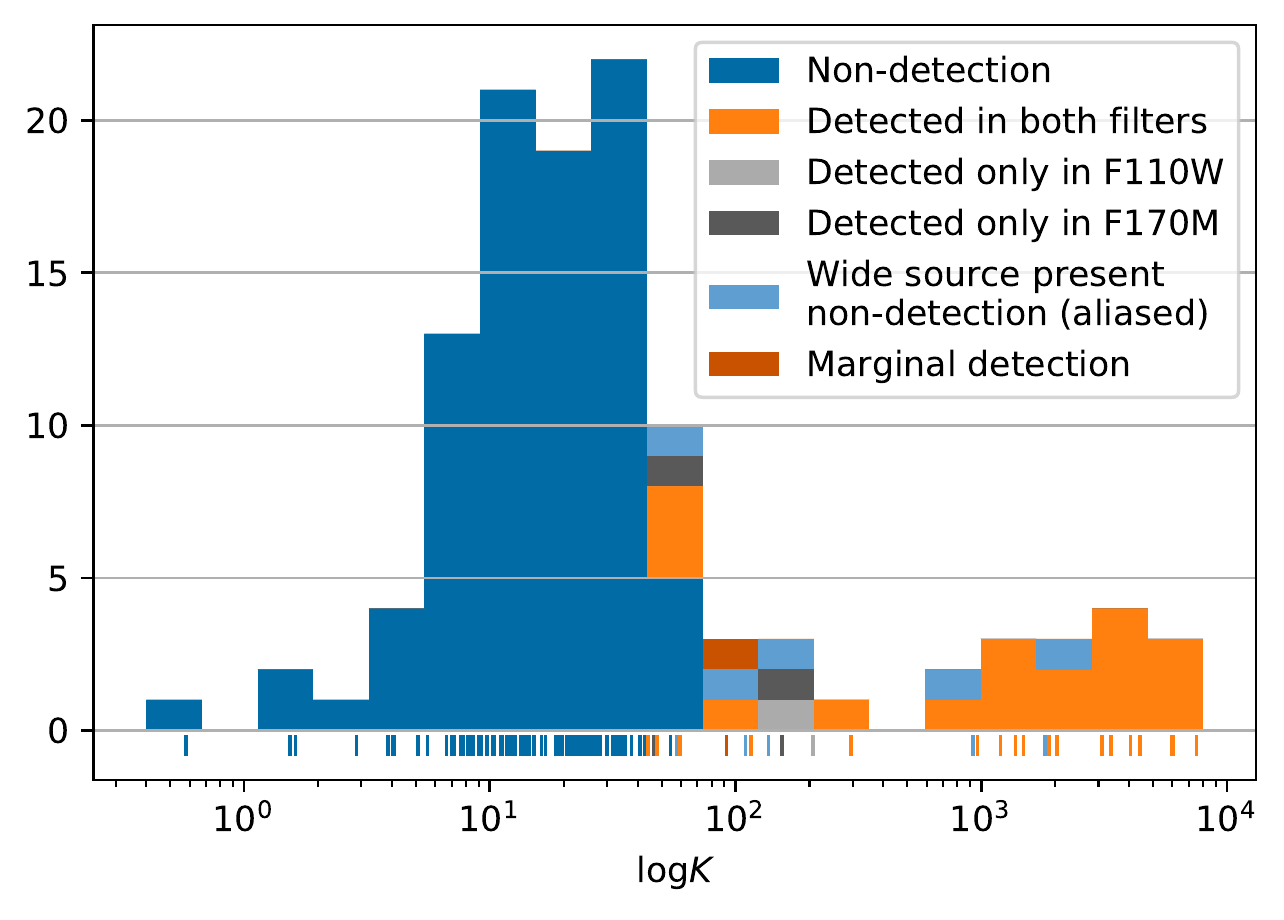}
    \caption{Histogram of Bayes factors (median value for the 5 calibrators) for the 117 observations in our sample, color coded by the assessment from our scramble and fit detection method. True (un-binned) positions are indicated with dashes under the histogram. Individual targets are discussed in Section~\ref{sec:scrambleVsBF}.} 
    \label{fig:BFhist}
\end{figure}

A histogram of Bayes factors (represented by the symbol $K$) for all of our targets is shown in Figure~\ref{fig:BFhist} color coded by our detection method described in Section~\ref{sec:detLim}. We distinguish between single targets, confident detections, confident detections in only one filter, a marginal detection, and undetected targets with wide separation companions (which KPI is insensitive to).

\citet{Jeffreys1961} set out guidelines for how to interpret $K$ ranging from $K<1$ supporting the single model to $K>100$ indicating ``decisive" evidence for a binary. This scale does not match well with our calculated values as all but one source strongly supports the binary model. While we do see a significant difference in $K$ between binaries and single sources, the differentiating value is much larger than \citet{Jeffreys1961} set out. This may be due to the large number of data-points we are fitting. 

A few sources which we designated as binaries had small $K$ values and a few which we designated as single had large $K$ values. These sources are discussed further in Section~\ref{sec:scrambleVsBF} and are known shortcomings of the available datasets and KP analysis in general. 

\section{Results}\label{sec:res}
We have detected companions in 21 observations corresponding to 19 binaries (two targets were observed twice) in our sample of 114 targets. We confirm one of the new kernel-phase detections presented in \citet{Pope2013}, and marginally recover a second, but none of their other ``confident" or ``marginal" detections. Even with our increased sensitivity at extremely close separations, we do not detect any new companions. Astrometry and photometry for the companions is presented in Table~\ref{tab:results}. Posteriors, kernel-phase correlation plots, and postage stamp images for each target are shown in Figure~\ref{fig:model} and the corresponding Figure Set. Our sample also includes 2 wide binaries which KPI is not sensitive to and thus we do not recover, as well as two observations of a suspected extremely tight binary which we do not recover a companion in (see Section~\ref{sec:nonDet}). 

\begin{deluxetable}{hlhCChCChCChCCh}
\tabletypesize{\scriptsize}
\tablecaption{Binary astrometry \& photometry\label{tab:results}}
\tablehead{\nocolhead{Program ID/Target ID} & \colhead{Source} & \nocolhead{Sep. [mas]} & \colhead{Sep. [mas]} & \colhead{$\sigma_\mathrm{sys, sep}$ [mas]} & \nocolhead{PA [deg]} & \colhead{PA [deg]} & \colhead{$\sigma_\mathrm{sys, PA}$ [deg]} & \nocolhead{F110W contrast} & \colhead{F110W contrast} & \colhead{$\sigma_\mathrm{sys, F110W}$} & \nocolhead{F170M contrast} & \colhead{F170M contrast} & \colhead{$\sigma_\mathrm{sys F170M}$} & \nocolhead{calibrators} }
\startdata
                  10143/U20004 &    2M 0004-4044 &        83.19 \pm 0.22 &        83.2 \pm 0.5 &                     0.12 &       44.34 \pm 0.26 &       44.3 \pm 0.6 &                        0.4 &                1.126 \pm 0.008 &                1.126 \pm 0.018 &           0.004 &                                    &                                    &                   &  2M 1539-0520, 2M 2148+4003, 2M 0727+1710, 2M 0355+1133, 2M 0243-2453 \\
                  10143/U13016 &    2M 0025+4759 &         334.5 \pm 0.4 &       334.5 \pm 0.9 &                     0.05 &      232.79 \pm 0.06 &    232.79 \pm 0.13 &                      0.005 &                1.353 \pm 0.015 &                1.353 \pm 0.033 &          0.0030 &                1.251 \pm 0.010 &                1.251 \pm 0.023 &          0.0010 &  2M 2139+0220, 2M 0243-2453, 2M 2351-2537, 2M 0228+1639, 2M 0825+2115 \\
                  10143/U20081 &    2M 0147-4954 &       138.85 \pm 0.20 &       138.8 \pm 0.4 &                      0.9 &       72.64 \pm 0.07 &     72.64 \pm 0.14 &                       0.20 &                2.345 \pm 0.015 &                2.345 \pm 0.030 &            0.22 &                2.022 \pm 0.008 &                2.022 \pm 0.016 &            0.08 &  2M 1058-1548, 2M 1421+1827, 2M 0123-4240, 2M 0926+5847, 2M 1750+1759 \\
       9833/SDSS-042348-041404 &    2M 0423-0414\tablenotemark{1} &       159.11 \pm 0.12 &     159.11 \pm 0.24 &                     0.15 &       19.77 \pm 0.04 &     19.77 \pm 0.08 &                       0.05 &                1.655 \pm 0.005 &                1.655 \pm 0.011 &           0.015 &                2.124 \pm 0.005 &                2.124 \pm 0.010 &           0.007 &                2M 0415-0935, 2M 2228-4310, 2M 1108+6830, 2M 1110+0116 \\
           11136/SDSS0423-04AB &    2M 0423-0414\tablenotemark{2} &        83.11 \pm 0.28 &        83.1 \pm 0.7 &                      0.8 &        183.1 \pm 0.4 &      183.1 \pm 0.8 &                        0.4 &                                    &                                    &                   &                  2.06 \pm 0.10 &                  2.06 \pm 0.23 &            0.25 &  2M 2228-4310, 2M 0415-0935, 2M 1624+0029, 2M 1217-0311, 2M 2339+1352 \\
                  10143/U10287 &    2M 0429-3123\tablenotemark{3} &         533.7 \pm 1.8 &           534 \pm 4 &                     0.17 &      285.16 \pm 0.17 &      285.2 \pm 0.4 &                      0.016 &                  3.50 \pm 0.10 &                  3.50 \pm 0.23 &           0.010 &                  2.77 \pm 0.06 &                  2.77 \pm 0.13 &           0.004 &  2M 1221+0257, 2M 0213+4444, 2M 1043+2225, 2M 0314+1603, 2M 2237+3922 \\
                  10143/U10617 &    2M 0700+3157 &       179.41 \pm 0.34 &       179.4 \pm 0.7 &                      0.8 &      105.68 \pm 0.09 &    105.68 \pm 0.18 &                       0.23 &                  4.43 \pm 0.06 &                  4.43 \pm 0.12 &            0.28 &                3.809 \pm 0.016 &                3.809 \pm 0.033 &            0.06 &  2M 1503+2525, 2M 1104+1959, 2M 0518-2828, 2M 1448+1031, 2M 0207+0000 \\
 9843/TWOMASSJ08503593+1057156 &    2M 0850+1057 &       132.24 \pm 0.30 &       132.2 \pm 0.7 &                      0.4 &      128.11 \pm 0.09 &    128.11 \pm 0.20 &                       0.16 &                2.929 \pm 0.022 &                  2.93 \pm 0.05 &           0.023 &                2.423 \pm 0.013 &                2.423 \pm 0.031 &           0.018 &  2M 1110+0116, 2M 0516-0445, 2M 0213+4444, 2M 1425-3650, 2M 2339+1352 \\
       9833/SDSS-092615+584721 &    2M 0926+5847 &        67.22 \pm 0.07 &      67.22 \pm 0.14 &                      0.4 &      134.26 \pm 0.14 &    134.26 \pm 0.28 &                        0.5 &                1.522 \pm 0.009 &                1.522 \pm 0.018 &           0.005 &                  2.70 \pm 0.12 &                  2.70 \pm 0.28 &            0.06 &                2M 1254-0122, 2M 0516-0445, 2M 1750+1759, 2M 0207+0000 \\
       9833/SDSS-102109-030420 &    2M 1021-0304 &       166.40 \pm 0.24 &       166.4 \pm 0.5 &                      1.4 &      244.83 \pm 0.07 &    244.83 \pm 0.13 &                       0.19 &                1.104 \pm 0.007 &                1.104 \pm 0.014 &           0.006 &                2.516 \pm 0.008 &                2.516 \pm 0.015 &            0.07 &                2M 2228-4310, 2M 1624+0029, 2M 2254+3123, 2M 0926+5847 \\
             11136/SDSS1534+16 &    2M 1534+1615 &         114.2 \pm 1.6 &           114 \pm 4 &                      1.3 &        310.6 \pm 0.7 &      310.6 \pm 1.6 &                        0.6 &                  2.96 \pm 0.15 &                  2.96 \pm 0.35 &            0.09 &                  2.55 \pm 0.12 &                  2.55 \pm 0.27 &            0.07 &  2M 0415-0935, 2M 2228-4310, 2M 1217-0311, 2M 1624+0029, 2M 2339+1352 \\
      9833/2MASS-155302+153237 &    2M 1553+1532 &       345.66 \pm 0.31 &       345.7 \pm 0.7 &                      0.7 &      189.69 \pm 0.06 &    189.69 \pm 0.13 &                       0.13 &                1.363 \pm 0.015 &                1.363 \pm 0.034 &           0.014 &                1.408 \pm 0.011 &                1.408 \pm 0.027 &           0.004 &  2M 0318-3421, 2M 2254+3123, 2M 0155+0950, 2M 0251-0352, 2M 2252-1730 \\
 9843/TWOMASSJ17281150+3948593 &    2M 1728+3948 &       157.75 \pm 0.15 &     157.75 \pm 0.28 &                     0.13 &       66.73 \pm 0.05 &     66.73 \pm 0.10 &                       0.05 &                1.390 \pm 0.009 &                1.390 \pm 0.018 &          0.0011 &                1.535 \pm 0.004 &                1.535 \pm 0.007 &          0.0031 &                2M 1428+5923, 2M 1421+1827, 2M 0830+4828, 2M 0207+0000 \\
             11136/SDSS2052-16 &  SDSS 2052-1609 &         102.1 \pm 0.4 &       102.1 \pm 0.9 &                     0.11 &         48.5 \pm 0.6 &       48.5 \pm 1.1 &                        0.6 &                  4.42 \pm 0.35 &                    4.4 \pm 0.7 &            0.14 &                  1.72 \pm 0.04 &                  1.72 \pm 0.09 &            0.04 &                2M 1624+0029, 2M 2228-4310, 2M 0415-0935, 2M 1110+0116 \\
                  10143/U20925 &    2M 2152+0937 &       254.24 \pm 0.27 &       254.2 \pm 0.6 &                    0.034 &       94.31 \pm 0.08 &     94.31 \pm 0.18 &                      0.009 &                1.157 \pm 0.008 &                1.157 \pm 0.019 &          0.0007 &                1.126 \pm 0.005 &                1.126 \pm 0.012 &          0.0005 &  2M 0624-4521, 2M 1721+3344, 2M 0911+7401, 2M 1300+1912, 2M 0443+0002 \\
                  10143/U20976 &    2M 2252-1730\tablenotemark{4} &       126.75 \pm 0.30 &       126.7 \pm 0.7 &                      0.9 &      353.96 \pm 0.08 &    353.96 \pm 0.17 &                        0.6 &                2.568 \pm 0.014 &                2.568 \pm 0.028 &           0.027 &                3.214 \pm 0.022 &                  3.21 \pm 0.04 &            0.32 &  2M 0251-0352, 2M 0155+0950, 2M 0423-0414, 2M 2254+3123, 2M 0415-0935 \\
               11136/2M2252-17 &    2M 2252-1730\tablenotemark{5} &          91.9 \pm 1.0 &        91.9 \pm 1.9 &                      2.0 &      172.89 \pm 0.30 &      172.9 \pm 0.6 &                        0.7 &                  5.82 \pm 0.34 &                    5.8 \pm 0.7 &            0.14 &                  3.29 \pm 0.20 &                    3.3 \pm 0.4 &             0.5 &                2M 2228-4310, 2M 0415-0935, 2M 1217-0311, 2M 1624+0029 \\
                  10879/U20979 &    2M 2255-5713 &       178.31 \pm 0.33 &       178.3 \pm 0.8 &                     0.27 &      172.64 \pm 0.09 &    172.64 \pm 0.22 &                       0.24 &                  5.16 \pm 0.06 &                  5.16 \pm 0.14 &            0.19 &                4.533 \pm 0.023 &                  4.53 \pm 0.06 &            0.10 &  2M 1110+0116, 2M 2339+1352, 2M 1217-0311, 2M 0908+5032, 2M 1043+2225 \\
                  10143/U12220 &    2M 2351-2537 &        62.65 \pm 0.18 &      62.65 \pm 0.34 &                      0.5 &      348.74 \pm 0.16 &    348.74 \pm 0.28 &                        0.8 &                  2.58 \pm 0.05 &                  2.58 \pm 0.09 &            0.13 &                  2.93 \pm 0.35 &                    2.9 \pm 0.9 &             0.7 &              SDSS 0837-0000, 2M 0825+2115, 2M 0228+1639, 2M 1213-0432 \\
\multicolumn{10}{l}{Marginal (see section \ref{sec:marginal})}\\
                  10879/U20866 &    2M 2028+0052 &          39.0 \pm 1.4 &        39.0 \pm 3.3 &                        4 &        118.3 \pm 0.5 &      118.3 \pm 1.0 &                          7 &                  1.71 \pm 0.29 &                    1.7 \pm 3.2 &             3.5 &                                    &                                    &                   &                2M 0911+7401, 2M 2036+1051, 2M 1439+1929, 2M 0500+0330 \\
\multicolumn{10}{l}{Spurious (see section \ref{sec:spurious})}\\
                  10879/U10018 &    2M 0024-0158 &             202 \pm 4 &           202 \pm 9 &                       10 &         33.4 \pm 0.6 &       33.4 \pm 1.3 &                        1.8 &                                    &                                    &                   &                      121 \pm 5 &                     121 \pm 11 &               8 &                2M 0523-1403, 2M 1155-3727, 2M 2002-0521, 2M 1721+3344 \\
          11136/DENIS0205-1159 &    2M 0205-1159 &       105.96 \pm 0.30 &       106.0 \pm 0.6 &                     0.07 &        349.8 \pm 0.5 &      349.8 \pm 1.0 &                       0.32 &                  1.75 \pm 0.04 &                  1.75 \pm 0.09 &           0.012 &                1.122 \pm 0.011 &                1.122 \pm 0.023 &           0.008 &                2M 2228-4310, 2M 0415-0935, 2M 1217-0311, 2M 2339+1352 \\
             11136/EPS-IND-BAB &    2M 2204-5646 &         127.9 \pm 1.1 &       127.9 \pm 2.3 &                     0.34 &        310.7 \pm 1.0 &      310.7 \pm 2.1 &                        1.8 &                  3.02 \pm 0.13 &                  3.02 \pm 0.28 &            0.14 &                  5.18 \pm 0.18 &                    5.2 \pm 0.4 &             0.4 &                2M 1624+0029, 2M 0415-0935, 2M 2228-4310, 2M 1254-0122 \\
\enddata
\tablecomments{1: Observations from program 9833 (7/2004) 2: Observations from program 11136 (8/2008) 3: 2M 0429-3123 is just wider separation than the grid used for detection limits but fitted parameters agree with literature values (see discussion of this source in Section~\ref{sec:knownBin}). 4: Observations from program 10143 (6/2005) 5: Observations from program 11136 (5/2008)}
\end{deluxetable}

\subsection{Known Binaries}\label{sec:knownBin}

We find that 12 sources agree well with previously published astrometry and photometry. This includes \textbf{2M 2351-2537}, one of the new extremely tight (62.7 mas, $0.4 \lambda/D$ in F110W and $0.66 \lambda/D$ in F170M) binaries discovered using KPI by \citet{Pope2013}. We discuss the remaining sources below, which have some discrepancy. No previously published astrometry and photometry could be found for the NICMOS imaging of \textbf{2M 1534+1615} from program 11136 \citep{Liu2007} though our fits visually agree with the position of the companion in the images. 

The companion to \textbf{2M 0004-4044} is not detected in the F170M filter. This target is a tight ($\sim 0.5\lambda/D$) equal brightness target which KPI is less sensitive to (since the system lacks asymmetry). The companion is however detected in the F110W filter where the diffraction limit is smaller. \citet{Pope2013} fit the full visibilities (rather than just the kernel phases) for this target to increase the sensitivity to the low-contrast companion. Our best fit parameters are slightly tighter and fainter than those of \citet{Pope2013}, perpendicular to the contrast-separation degeneracy seen in tight-separation fits. 

\textbf{2M 0025+4759} is slightly wider than the best fit parameters from \citet{Pope2013}. Our contrast in F110W is consistent but our F170M contrast ($C_\mathrm{F170M}=1.25\pm0.02$) is slightly higher than both \citet{Pope2013} ($C_\mathrm{F170M}=1.03\pm0.04$) and \citet{Reid2006} ($C_\mathrm{F170M}=1.11$). \citet{Pope2013} also fit the visibilities of this roughly equal brightness source rather than the kernel phases since they knew it was a low-contrast companion.
  
\textbf{2M 0423-0414} was observed by both \citet{Burgasser2006} and \citet{Liu2007}. We detect the companion in F170M in both epochs but in F110W in only one epoch. The companion is much tighter in the second epoch and the observations are much lower SNR so it is not surprising that we do not detect it in the shorter wavelength filter. Our astrometry and photometry is consistent with that presented by \citet{Dupuy2017} though we do pick up some separation-contrast degeneracy in the later, close-separation epoch (our separation is slightly tighter and contrast slightly higher). \citet{Dupuy2017} also derived a dynamical mass for this system from its orbital motion in these and other \emph{HST/ACS} observations. 

\textbf{2M 0429-3123} is just wider than the upper edge of our detection limit grid but the best fit parameters are consistent between calibrators and agree with literature values. We thus consider this a significant detection. Our best fit separation ($534\pm4$~mas) splits the difference in separation between \citet{Reid2006} (550~mas) and \citet{Pope2013} ($525.2\pm1.2$~mas) while our contrast is consistent. 
  
\textbf{2M 0926+5847} is consistent with \citet{Burgasser2006} with the exception of F170M (they report $C_\mathrm{F170M}=1.4\pm0.4$ while we report $C_\mathrm{F170M}=2.70\pm0.28$). Our F170M contrast value is more consistent with \citet[][$C_\mathrm{F170}=2.2\pm0.2$]{Dupuy2017}, though still slightly higher contrast. KPI is much more precise than PSF fitting at the separation of this companion ($67.22 \pm 0.14$~mas) and our contrast values are consistent with the color/spectral type relations of T dwarfs \citep{Leggett2002}. 

\textbf{SDSS 2052-1609} was observed at low SNR in program 11136 \citep{Liu2007}. Our separation and position angle agree with \citet{Dupuy2017} with higher precision though our best fit F110W contrast ($C_\mathrm{F110W}=4.4\pm0.7$) is significantly higher than theirs ($C_\mathrm{F110W}=1.2\pm0.4$). Our high contrast F110W fit does not seem consistent with the images even though this target should be easily recovered by KPI. It is possible that the super-Gaussian window we used to exclude noise far away from the target is too wide ($\sigma=25\lambda/D$) for these low SNR observations where the wings of the PSF are not present. Orbital motion is seen between this and other observations and was fit by \citet{Dupuy2017}.

\textbf{2M 2252-1730} was observed by \citet{Reid2006} and later by \citet{Liu2007}. Our best fit parameters are consistent with the \citet{Reid2006} and \citet{Pope2013} epoch while we pick up some separation-contrast degeneracy in the later \citet{Liu2007} epoch. Our values (sep~$=91.9\pm1.9$~mas, $C_\mathrm{F110W}=5.8\pm0.7$) are slightly wider and significantly higher contrast than those presented by \citet{Dupuy2017} (sep~$=87\pm3$mas, $C_\mathrm{F110W}=1.9\pm0.7$). As with \textbf{SDSS 2052-1609}, our high contrast F110W fit is likely caused by too wide of a super-Gaussian window. Orbital motion is seen between the two epochs and was fit by \citet{Dupuy2017}. 
 
\subsection{Marginal Detection}\label{sec:marginal}

\textbf{2M 2028+0052} is not detected in F170M and just barely fails to pass our confidence test in F110W. While four of five calibrators produce fits which are well grouped, one pair of calibrators is inconsistent with each other in position at $5.16\sigma$ (just above our $5\sigma$ cutoff) and another pair is inconsistent with each other in contrast at $1.07\sigma$ (just above our $1\sigma$ cutoff). This target also has a slightly elevated Bayes Factor with respect to sources we categorize as single. We thus present this as a marginal detection. Adding more calibrators and a more careful treatment of bad pixels, thus reducing the noise, will help to refine this determination. It also has an elevated RUWE \citep[1.418;][]{Gaia2020}, even when fitting for the astrometric color \citep{Lindegren2021}, suggesting the presence of a companion. \citet{Dahn2017} also notes elevated and periodic ($\lesssim2$years) residuals in their parallax solution.

\begin{figure}
\centering
\includegraphics[width=\textwidth]{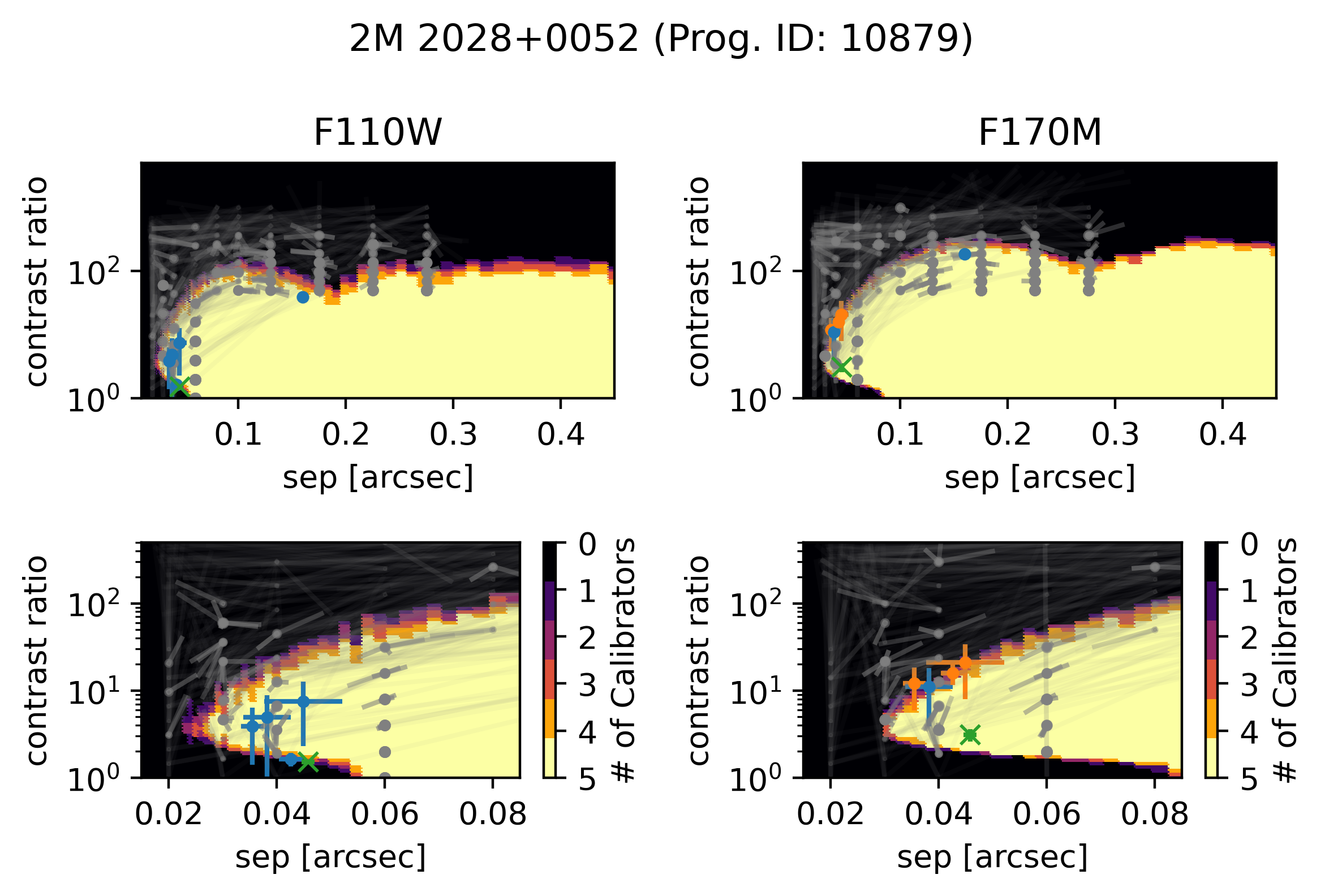}
\caption{Similar to Figure~\ref{fig:nonSigDetLim} but for the marginal detection of 2M 2028+0052. The best fit parameters from \citet{Pope2013} are indicated by the green X. Similar plots for the spurious detections are shown in Figure Set 9. The full figure set (4 images) is available online at \citet{figSets}.}
\label{fig:margDetLim}
\end{figure}

\begin{figure}
\includegraphics[width=\textwidth]{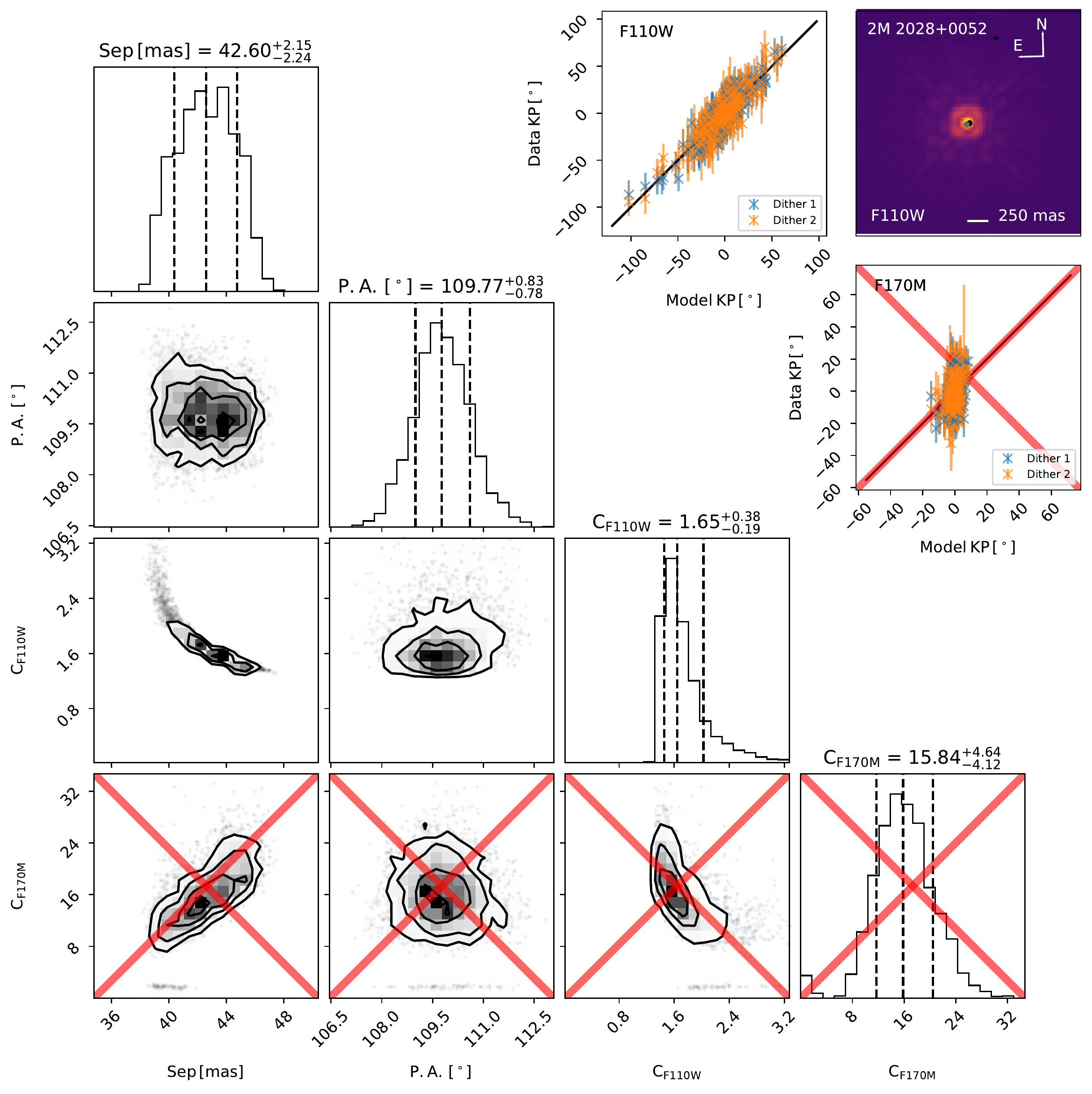}
\caption{Similar to Figure~\ref{fig:model} but for the marginal detection of 2MASS 2028+0052. Red X's indicate no detection in F170M.}\label{fig:model2028}
\end{figure}

Our best fit parameters are consistent with those of \citet{Pope2013}, when considering the intercalibrator systematic error-bars. Our best fit parameters are heavily weighted toward a single higher precision fit, while the fits using the other calibrators are much lower precision (as can be seen in the detection limit plot shown in Figure~\ref{fig:margDetLim}. Figure~\ref{fig:model2028} shows the corner plot for the high precision fit. There appears to be a small bump in the posterior at a lower contrast in F170M which may be consistent with the \citet{Pope2013} contrast.

%
%
%
%

\subsection{Spurious Detections}\label{sec:spurious}

\begin{figure}
\includegraphics[width=\textwidth]{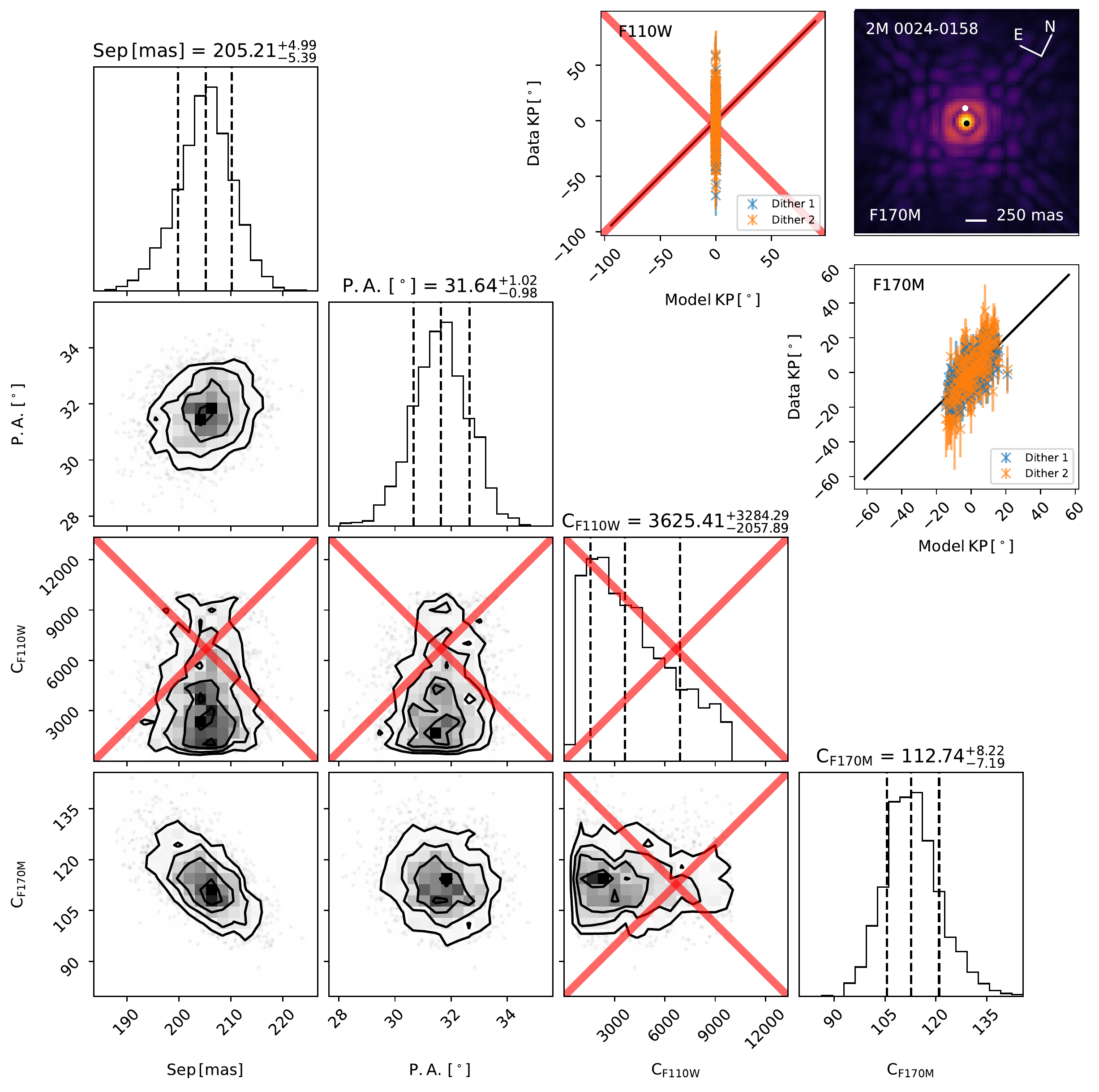}
\caption{Similar to Figure~\ref{fig:model} but for 2MASS 0024-0158. Large red X's over the 1- and 2D histograms indicate no detection in the F110W filter. This fit locked onto the signal produced by a bad pixel.}\label{fig:model0024}
\end{figure}

\begin{figure}
\includegraphics[width=\textwidth]{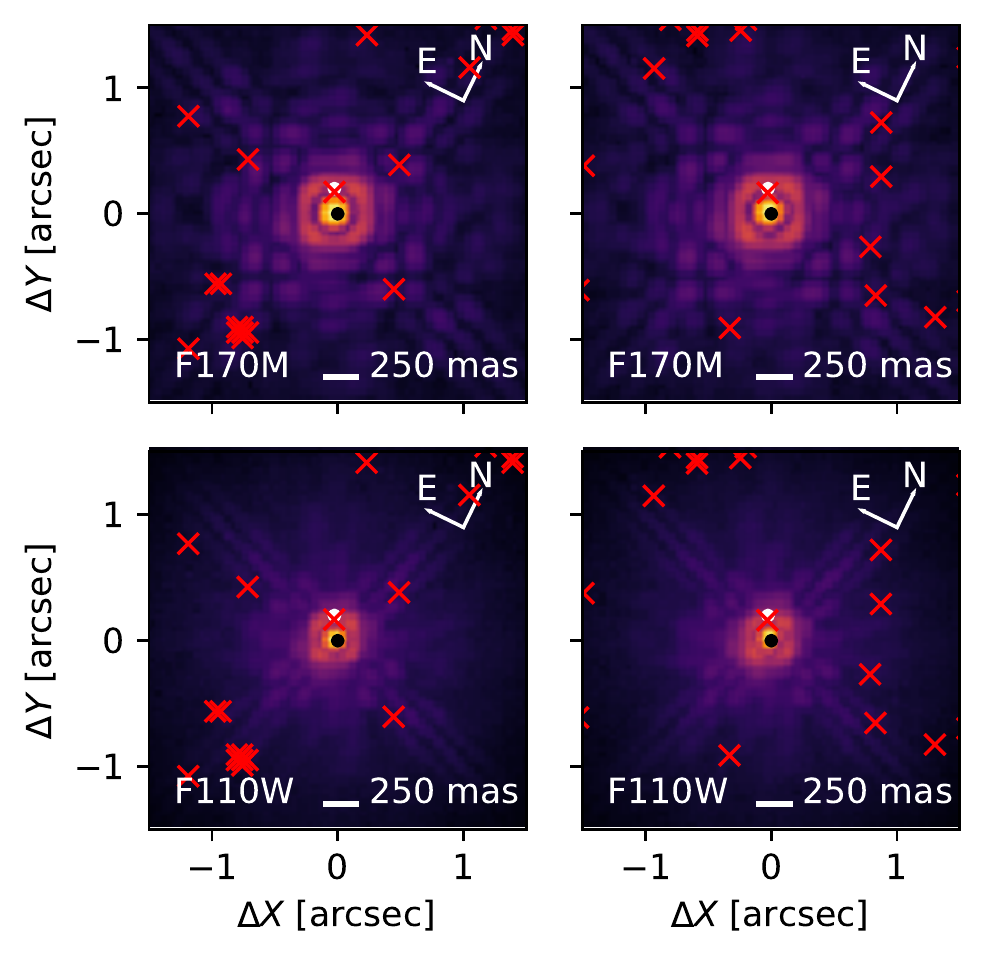}
\caption{Both filter and dither position images of 2M~0024-0158. The location of the primary is indicated by a black circle and the location of the best fit ``companion" is shown by a white circle. The locations of bad pixels are indicated with red X's. Note that in both dither positions there is a bad pixel at almost exactly the same location relative to the target and that in F170M the bad pixel is located in the first dark ring while in F110W it is located in the first bright ring. Our bad-pixel rejection imperfectly replaced the bad pixel on the dark ring and yielded a spurious faint detection in F170M.}\label{fig:im0024}
\end{figure}

We detect a high contrast ``companion" around \textbf{2M~0024-0158} in F170M but with no detection in F110W. Figure~\ref{fig:model0024} shows the corner plot, kernel-phase correlation plots, and a representative image for this target. Visual inspection of the images and associated data-quality frames demonstrates that this detection is spurious. The position of the detection almost exactly lines up with bad pixels which were coincidentally located at the same position, relative to the target, in both dither positions. Our bad-pixel correction technique (replacing the pixel with the median of the 8 surrounding pixels) injected a brighter than expected pixel in the F170M images, since it fell in the first dark airy ring. This pixel was then interpreted as a companion. In the F110W images the bad pixel fell in the first bright airy ring and did not inject the signal of a companion. This can be seen in Figure~\ref{fig:im0024}. We present this case as a cautionary tale to show how image-plane based bad-pixel rejection routines can inject signals in the Fourier plane. In the future, bad-pixel rejection techniques in the Fourier plane should be used as described by \citet{Ireland2013} and implemented in \citet{Kammerer2019}. 

\begin{figure}
\includegraphics[width=\textwidth]{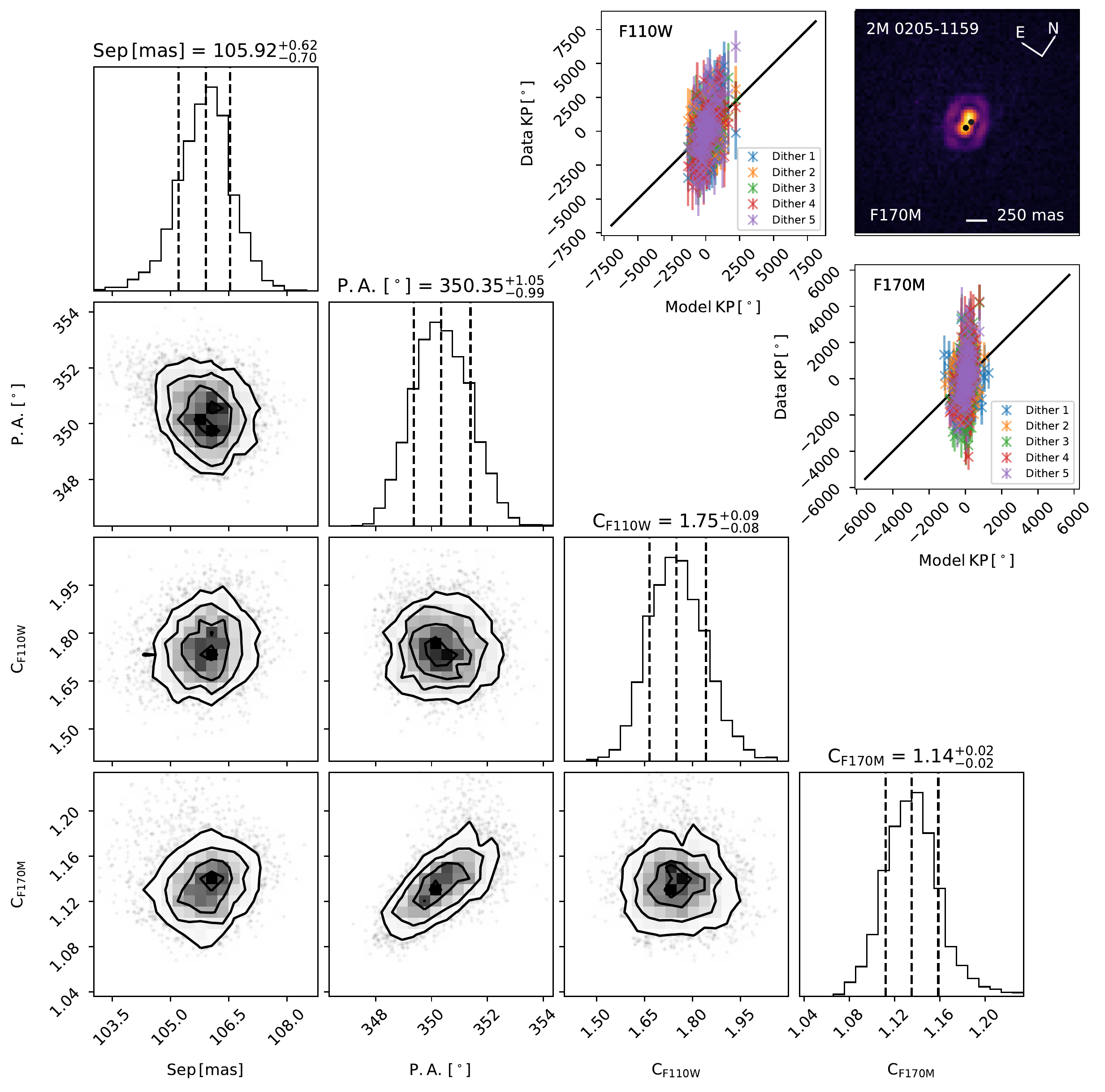}
\caption{Similar to Figure~\ref{fig:model} but for the spurious fit to 2MASS 0205-1159. This source is a triple system and thus is not well fit with a binary model. }\label{fig:model0205}
\end{figure}

\begin{figure}
\includegraphics[width=\textwidth]{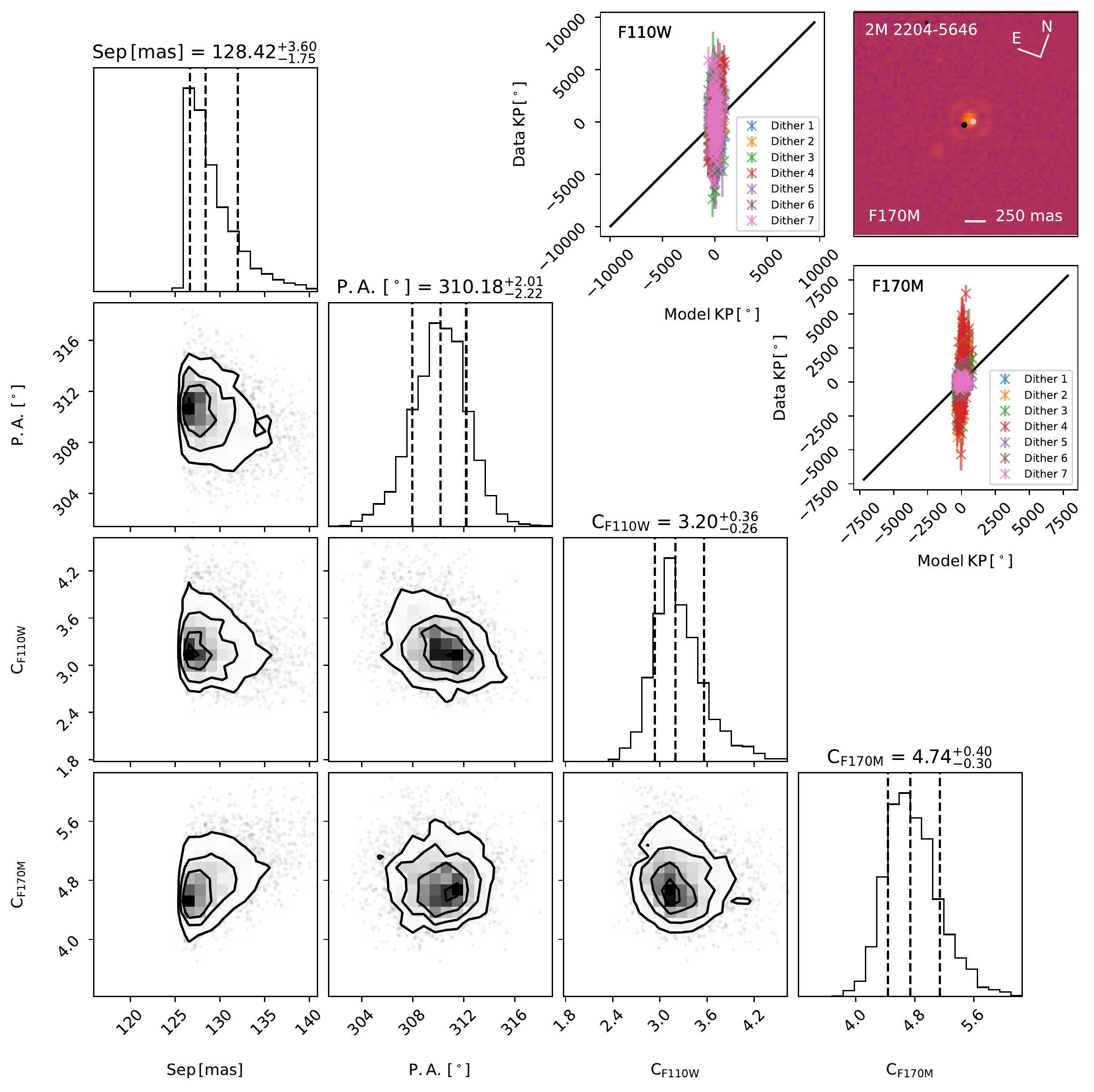}
\caption{Similar to Figure~\ref{fig:model} but for 2MASS 2204-5646. Our binary model did not recover the true values of the wide companion in this low SNR observation. }\label{fig:model2204}
\end{figure}

A simple visual inspection of the fits to \textbf{2M~0205-1159} and \textbf{2M~2204-5646}, shown in Figures~\ref{fig:model0205} and \ref{fig:model2204}, shows a clear discrepancy between the visually obvious companion and the best fit parameters. Both of these targets were observed in program 11136 which observed using a large number of filters, sacrificing SNR for more wavelength coverage. Therefore, while the core of the PSFs are clear, little to none of the wings of the diffraction pattern is recovered. While we present our best fit parameters for completeness of the survey, we will give these sources a more careful treatment in a future letter. In addition, \textbf{2M 0205-1159} is a previously known triple system \citep{Bouy2005}. The position and contrast of our best fit companion is likely a combination of the B and C components. Fitting this system with a triple point source model will be done in a future letter as the addition of a third source adds nontrivial degeneracies in the model. 

\subsection{Notable Non-detections}\label{sec:nonDet}

We do not significantly detect any of the marginal detections presented in \citet{Pope2013} and only detect two of their five confident detections (\textbf{2M~2351-2537} significantly and \textbf{2M~2028+0052} marginally, discussed above). While some of our best fit parameters with a \emph{single} calibrator are consistent with their results and that single fit may be significant (e.g. \textbf{2M~0314+1603} and \textbf{2M~1936-5502}), they are not consistent between multiple calibrators (see Figure~\ref{fig:noteNonDetLim} and the relevant figure in Figure Set 7) and we therefore do not classify them as detections. This demonstrates the strength of and need for a robust calibration strategy. Due to the extreme separation of these proposed companions, a ground-truth validation of these companions is difficult though not impossible. Possible techniques include LGSAO on a large ground based telescope or astrometry \citep[e.g. future work of][]{Gaia2021,Dahn2017}. While binary fits are still not part of Gaia data releases, RUWE can be used as a proxy for the presence of a companion \citep[][Kraus et al. in prep.]{Lindegren2018,Belokurov2020,Lindegren2021}. All of the marginal and confident detections (with the exception of the two sources we recover) have RUWE of $\sim1$ \citep{Gaia2020}. If these companions are real we would expect them to have an elevated RUWE like \textbf{2M 2351-2537} ($\mathrm{RUWE}=5.417$), though equal brightness companions have less of an astrometric signal and faint targets have inherently more noise.

It is also difficult for our implementation of KPI to recover wide companions at separations greater than $\sim0.6$ arcsec accurately. These companions suffer from aliasing in the Fourier plane as sampled by our aperture model and do not require high resolution where KPI excels. Since these companions are generally noticeable under visual inspection, one could tune the prior to isolate the true signal from aliases but it would be more straightforward to use classical PSF fitting techniques. 

\textbf{2M 0915+0422} is a wide separation companion \citep[$738.6\pm0.15$~mas][]{Pope2013} and thus our pipeline is strongly affected by aliasing in the Fourier plane. Our fits do not recover the literature values for this companion. We do detect a significant companion but at an alias of the known companion. 

\textbf{2M 1707-0558} is also a wide separation companion \citep[$1009.5\pm1$~mas;][]{Pope2013} and again our pipeline is strongly affected by aliasing in the Fourier plane. We do recover the companion in two of five calibrators but the aliasing is strongly affecting our best fit parameters so we do not present them here. Our best fit separation is slightly wider than both \citet{Reid2006} and \citet{Pope2013} though our contrasts are more consistent with \citet{Reid2006}. 

\textbf{2M 1705+0516} and \textbf{2M 1731+2721} both have wide objects in the images that, based on color and follow up observations (Dupuy, private communication), appear to be background sources. 

We do not detect a companion to \textbf{2M 0518-2828} in either of the two epochs (programs 10247 and 11136). \citet{Cruz2004} published the near infrared spectrum of this object which had features of both an L and T dwarf, suggesting that it was an unresolved binary. \citet{Burgasser2006} then published \emph{HST/NICMOS} images of this target and found that the PSF was ``slightly elongated", more so in the shorter wavelength filters. The elongation is extremely small in F110W and absent in F170M so our non-detection is not unexpected. \citet{Burgasser2006} measured a separation of $51\pm12$~mas and a contrast of $C_\mathrm{F110W}=2.1\pm1.0$ ($0.8\pm0.5$~mag) and $C_\mathrm{F170M}=2.3\pm1.3$ ($0.9\pm0.6$~mag). The separation is what is limiting us in this case as, for context, it is roughly 10 mas tighter than \textbf{2M 2351-2537} at slightly lower contrast. If KPI can be done in F090M (larger pixel scale relative to $\lambda/D$ and larger aberrations relative to $\lambda$) a detection may be possible. 

\subsection{Survey Detection Limits}\label{sec:surveyDetLim}
\begin{figure}
\includegraphics[width=0.5\textwidth]{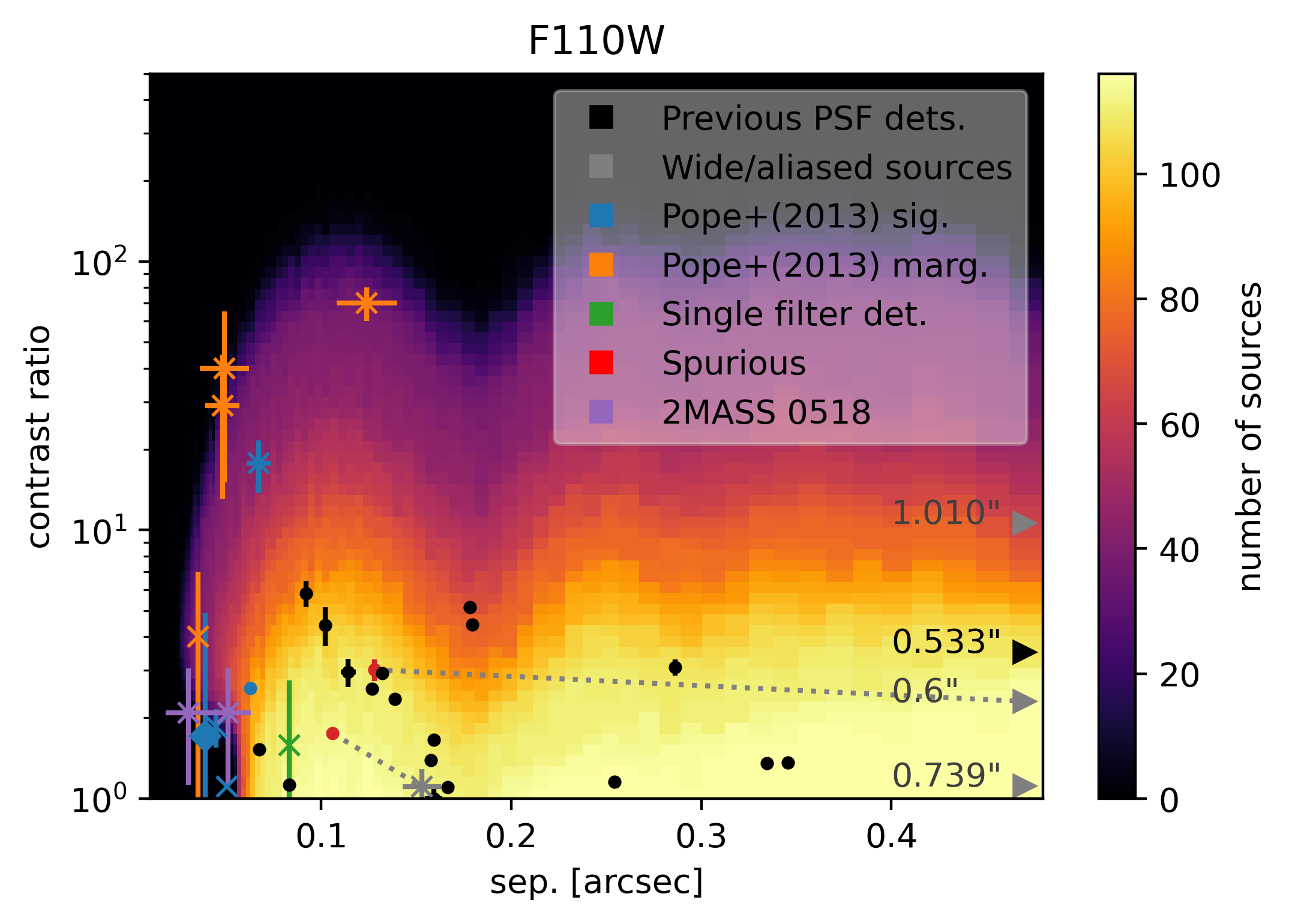}
\includegraphics[width=0.5\textwidth]{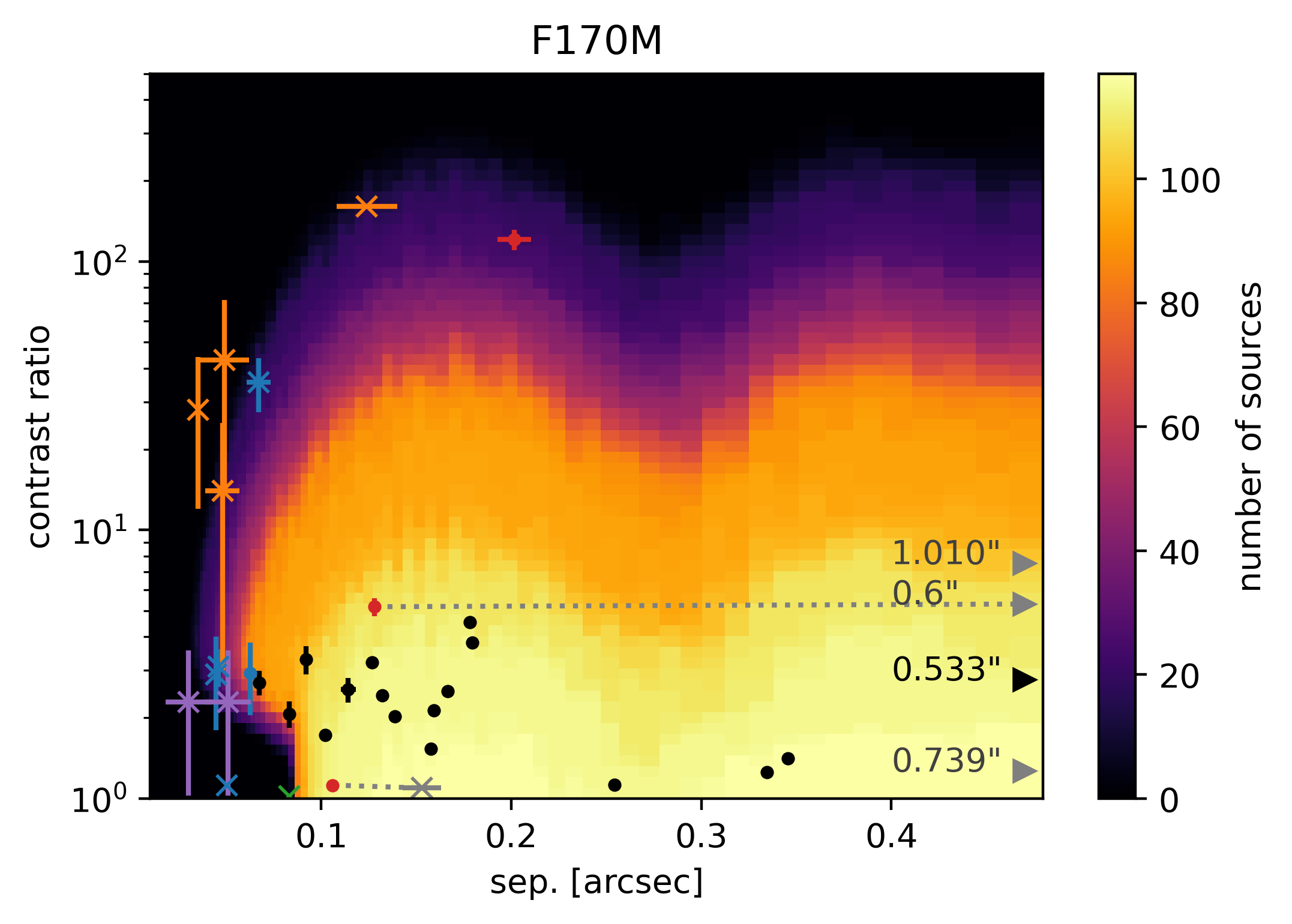}
\caption{Stacked $5\sigma$ detection limits as a function of separation for all sources in the survey. Filled circles indicate the positions of detected companions, the diamond is our one marginal detection, and X's indicate literature values for non-detections. Points are color coded by type of detection and are outlined in the legend. Black symbols indicate companions previously found by PSF fitting. Blue and orange symbols indicate the ``significant" and ``marginal" detections from \citet{Pope2013}, respectively. Green X's indicate the literature contrast values of the two sources (one in each filter) which we only detected in one filter. Red points indicate the positions of ``spurious" detections (see Section~\ref{sec:spurious}) with dotted lines to gray X's indicating the literature values. Purple symbols indicate the position of 2MASS 0518-2828 (see Section~\ref{sec:nonDet}). Two sources were extremely wide and were not recovered due to strong aliasing and are shown by gray arrows, at a separation given by the labeled text. The survey detection limits in both filters, in units of number of sources, are available as the Data behind the Figure online at \citet{figSets} or in the .tar.gz file available on arXiv (see README.txt for info on the .fits files). }\label{fig:allDet}
\end{figure}

A superposition of all detection limits is shown in Figure~\ref{fig:allDet}. Detection limits for each target are shown in Figures~\ref{fig:nonSigDetLim}, \ref{fig:sigDetLim}, and \ref{fig:noteNonDetLim} and the corresponding Figure Sets, and are summarized in Tables~\ref{tab:detLim1}--\ref{tab:detLim3} at a set of separations. In the best cases, significant detections of companions can be achieved up to a contrast of $\sim100:1$ and down to a separation of $\sim0\farcs1$. Below $\sim0\farcs1$ the contrast limit drops steeply and becomes bimodal, with KPI becoming insensitive to equal-brightness companions at extremely tight separations due to the loss of asymmetry. A full treatment of the covariance matrix could increase the contrast limits at large separations due to the consideration of photon noise \citep{Ireland2013}.

\begin{figure}
    \centering
    \includegraphics[width=\textwidth]{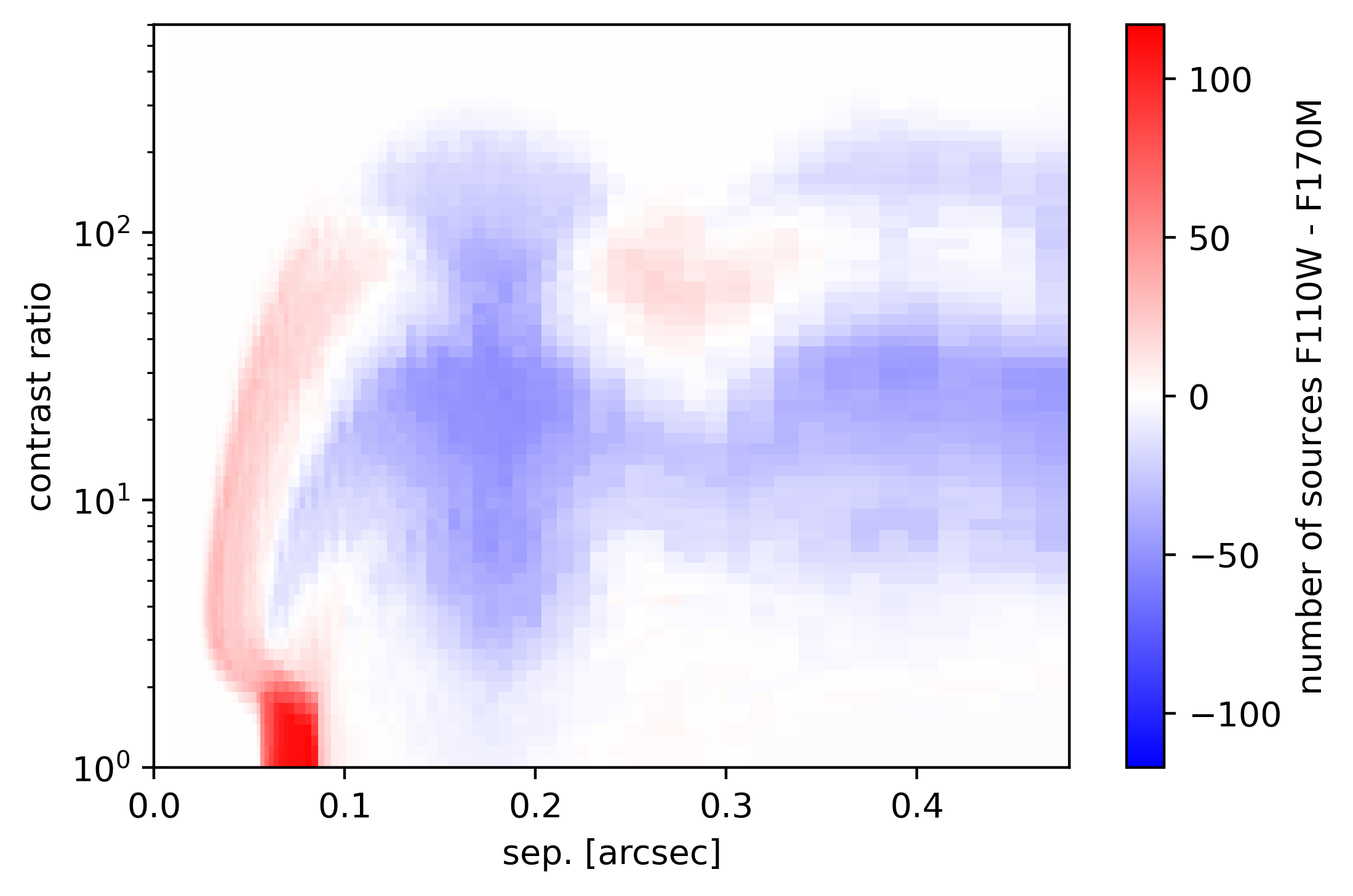}
    \caption{The difference in sensitivity between F110W and F170M (given in number of observations our analysis was sensitive to out of a total of 117) as a function of separation and contrast. F110W is more sensitive at close separations where the diffraction limit dominates, while F170M is more sensitive globally since aberrations are smaller with respect to the longer wavelength. }
    \label{fig:filterDiff}
\end{figure}

Comparing the difference in sensitivity between the two filters, as shown in Figure~\ref{fig:filterDiff}, reveals some clear takeaways. At close separation the shorter wavelength F110W filter is more sensitive since the diffraction limit is smaller. At wider separations observations in the longer wavelength F170M filter are more sensitive since optical aberrations are smaller with respect to the wavelength of light. These two filters also work in harmony by canceling out the dips in sensitivity as the brightness of the airy pattern fluctuates. A detection in two filters is also helpful to discern between real detections and spurious detections (such as for \textbf{2M 0024-00158}, caused by a bad pixel).

\section{Discussion}\label{sec:disc}

\subsection{Binary Frequency}\label{sec:binFreq}
While we present astrometry and photometry for all detections, we must restrict our analysis of binary populations to targets in programs 9833 \citep{Burgasser2006}, 10143 \citep{Reid2006}, and 10879 \citep{Reid2008} and we must exclude programs specifically targeting binaries (9843, 10247, 11136). This removes any bias in our sample toward binarity and enables a study of binary demographics. This brings the total number of objects in our sample down to 106 (from 114). The number of binary systems in our statistical sample drops to 15 (from 21) when also accounting for the 2 wide companions which KPI is not sensitive to.

To compare our binary fraction with \citet{Reid2008} and \citet{Pope2013} we restrict our sample to 20~pc. This is much more straightforward than in the previous studies due to recent parallax surveys \citep[e.g.][]{Gaia2018,Gaia2021,Best2020}. Our total number of sources then drops to 83, including 10 binaries while excluding 5. If we include \textbf{2M 0036+1821} and \textbf{2M 0355+1133}, which we do not detect companions around but were observed to have companions by \citet{Bernat2010}, this number rises to $12/83$. Estimating the uncertainty using binomial statistics \citep{Burgasser2003b} gives a binary fraction of $\epsilon_b = 14.4^{+4.7}_{-3.0}$\%. This binary fraction is slightly lower than the number presented in \citet{Pope2013} ($\epsilon_b = 17.2^{+5.7}_{-3.7}$\%) though they are consistent within the quoted errors. 

The typical mass of a brown dwarf is significantly lower than the Jeans mass of a collapsing protostellar cloud. Thus two theories for the formation mechanism of brown dwarfs have been proposed: gravoturbulent collapse and embryo ejection \citep{Basu2012}. Since embryo ejection is an energetic event, the binary fraction it produces is small: $<5-8\%$ \citep{Bate2002,Bate2012}. Our significantly higher binary fraction, and that of \citet{Pope2013} and other observational studies \citep[e.g.][]{Reid2006,Reid2008}, supports the conclusion that embryo ejection plays a minimal role in brown dwarf formation. A more detailed demographic analysis of the full unbiased sample (not distance limited but still excluding programs targeting binaries) will be presented in a future work using a Bayesian population analysis similar to \citet{Allen2007}. The semimajor-axis and mass-ratio distributions will be important to quantify in establishing the companion frequency, especially since all of these quantities are covariant when considering typical detection limits.

\subsection{Implications for Dynamical Mass Measurements}\label{sec:dynMass}
Our sample includes three targets which were observed twice. We detect companions around two of these sources and see orbital motion between the two epochs. Our astrometry for these two sources and six other sources with only one epoch is consistent with \citet{Dupuy2017} though our fits are much higher precision. Two other sources are listed in Table 3 of \citet{Dupuy2017} which have \emph{HST/NICMOS} imaging, though in filters other than F110W and F170M. These sources, along with others with relevant imaging in other filters will be analyzed in a future letter. Since no specific observing set up is required for KPI (unlike NRM observations), KPI is a particularly powerful tool for resolving binary orbits since it allows precovery of relative astrometry and photometry from archival observations of known binaries even when the binary was undetected or had much larger uncertainties when studied with image-plane analysis techniques.

\subsection{Scramble and fit detection limits vs. Bayes factors}\label{sec:scrambleVsBF}

One caveat of our scramble and fit method of measuring detection limits is that it does not recover the deepest possible contrast limit when the target has a strong binary signal. When the kernel phases are scrambled to generate a new realization of the noise, we want no signal to be present, so we subtract off the best fit model. In this survey where we treated all sources the same, we simply subtracted the best fit single point source model to remove any small position offset. When a binary is present, subtracting off only a signal source leaves signal behind, raising the measured noise threshold. To properly characterize the sensitivity around a non-single source, the proper model (with the correct parameter values) must be subtracted. This is particularly difficult when triple sources are possible (e.g. \textbf{2M 0205-1159}). For this source, our fitting routine failed to produce the companion fit that appears to match under visual inspection (because it is a triple) though it was characterized as significant. On the other hand, Bayes Factors are agnostic of the best fit parameters as they consider the entire volume of parameter space. They are however difficult to interpret if there are unmodeled systematics. For example, a few wide binaries which are aliased and thus difficult to recover with KPI are not considered significant detections by our scramble-and-fit method but do show high Bayes factors, indicating that a binary model is preferred even if our best fit parameters are clearly incorrect. 

As seen in Figure~\ref{fig:BFhist}, there are a few sources with high Bayes factors which did not pass our criteria for a detection and \emph{visa versa}. The three sources with the highest Bayes factors (which suggest binarity), but for which we do not find a successful KPI fit, are \textbf{2M 0429-3123}, \textbf{2M 1707-0558}, and \textbf{2M 0915+0422} (in order by descending Bayes factor). These are all wide binaries that are aliased by our aperture model and KPI struggles to accurately recover. We do characterize \textbf{2M 0429-3123} as a binary as its best fit parameters match with literature values while the companion was just outside our grid for detection limits. On the other hand our fits to \textbf{2M 1707-0558}, and \textbf{2M 0915+0422} both picked up on aliases and failed to recover the visually obvious companion. Similarly \textbf{2M 1705-0516} and \textbf{2M 1731+2721} both have high Bayes factors and a wide (aliased) secondary source in the images that is a background source. A more careful treatment of the Bayesian prior could encourage the fits to converge on the non-aliased signal, though it is clear from the Bayes factor that these sources prefer the binary model over the single one. Since Bayes factors are independent of best fit parameters, they allow us to infer the presence of a companion even if our best fit parameters are clearly aliased and thus incorrect.

Our marginal detection of \textbf{2M 2028+0052} is also supported by its elevated Bayes factor with respect to other single sources. This source is discussed further in Section \ref{sec:marginal}. 

Conversely the companions which we characterized as detections which had the lowest Bayes factors include \textbf{2M 0205-1159} and \textbf{2M 2204-5646}, two of our ``spurious" detections (one is a low SNR wide binary and the other is a triple). Detections with relatively low SNR observations, including \textbf{2M 1534+1615}, \textbf{SDSS 2052-1609}, and \textbf{2M 2252-1730}, also had low Bayes factors. In fact, all of these targets were observed in program 11136, which traded exposure time in the main filters for wider wavelength coverage in more filters. This indicates that high SNR observations, which significantly recover the wings of the PSF, are key to confident and accurate KPI detections. 

We have also examined a modified version of the likelihood ratio test discussed in \citet{Ceau2019}. Instead of the null hypothesis being all 0 kernel-phases we have replaced it with our single source model, allowing for a small centroid offset. This test gives similar results to the Bayes Factors though with slightly more overlap between significant detections and non-detections.

\subsection{KPI In Context}\label{sec:KPcontext}

Since our detection significance is determined on a source-specific basis, our survey wide stacked detection limit is as close as we can come to a pipeline-specific contrast curve similar to Figure~2 and 3 of \citet{Pope2013} (calculated using an injection-recovery grid with a noise term representative of the sample as a whole). We also use a $5\sigma$ (99.99994\%) detection threshold and separate out the two filters while \citet{Pope2013} presents only a single 99.9\% contrast curve. With the aforementioned caveats, the detection limits of our two KPI surveys are fairly similar. Both show a cutoff to equal-brightness companions at around 55~mas (in F110W for our survey, our cut-off in F170M is closer to 80~mas) with slightly higher contrast detections possible down to roughly 30~mas. Our high-contrast limits at wider separations are more difficult to compare given the difference in confidence threshold but appear qualitatively similar, with high confidence detections up to contrast values of roughly 100:1.

Comparing our KPI contrast limits to imaging surveys \citep[e.g. ][]{Reid2006,Reid2008} shows the power of KPI at close separations. Imaging sensitivity significantly drops at separations below $\sim0\farcs5$ while ours is flat down to less than $\sim0\farcs1$. At wide separations KPI suffers from aliasing in the Fourier plane and is ineffective at detecting sources without prior knowledge. In this wide separation domain PSF fitting performs well and KPI is not necessary. \citet{Sallum2019} showed, using simulated \emph{JWST} observations that NRM/AMI outperforms KPI at separations below the diffraction limit by 0.5--1 mag while KPI is comparable to or better than NRM/AMI outside the diffraction limit.

\section{Conclusions}\label{sec:conc}
We have presented a new pipeline named Argus for super-resolution detection of companions using KPI. The pipeline generates, calibrates, and fits kernel phases from high resolution images. The pipeline uses a scramble and fit method (similar to previous NRM pipelines) to determine the detection significance, though it can also use Bayes factors to determine if the source is non-singular. Our multi-calibrator method for determining if a detection is real is much more strict than previous KPI pipelines, though it finds roughly similar contrast limits. The pipeline is open source and is available on GitHub (see Footnote~\ref{footnote}).

We have demonstrated the pipeline on the entire \emph{HST/NICMOS} F110W and F170M image archive of nearby brown dwarfs (observed in 7 different programs). We recover 19 known companions (including two targets with two epochs) from our sample of 114 targets. We confirm one of the four new companions discovered by \citet{Pope2013} and marginally recovered a second, but we recover none of their marginal detections. We conclude that robust calibration with multiple well-matched calibrators is critical for determining the validity of a candidate companion, especially in the super-resolution regime where no spatially distinct PSF is present. 

We report a binary fraction of $\epsilon_b=14.4^{+4.7}_{-3.0}$\%, consistent with previous studies. We will perform a more detailed population level analysis of the catalogue presented here in a future paper. This work will simultaneously characterize the binary fraction, semimajor axis distribution, and mass ratio distributions using a Bayesian framework similar to \citet{Allen2007} to better understand how these low mass objects form. Future upgrades to the pipeline could include more robust bad pixel rejection techniques \citep{Ireland2013} and triple point source models. It also should be straightforward to make the code easily adaptable to other space based cameras amenable to KPI such as the high resolution channel of \emph{HST/ACS} and the imaging instruments aboard \emph{JWST}.

\acknowledgments
We thank Trent Dupuy, Will Best, and Michael Ireland for useful discussions about this work, and many others who have offered their thoughts at conferences. We also thank the anonymous referee for their helpful feedback which improved the manuscript. This work was funded by \emph{HST} program AR-14561. This work has benefited from The UltracoolSheet, maintained by Will Best, Trent Dupuy, Michael Liu, Rob Siverd, and Zhoujian Zhang, and developed from compilations by \citet{Dupuy2012}, \citet{Dupuy2013}, \citet{Liu2016}, \citet{Best2018}, and \citet{Best2021}. This work has made use of data from the European Space Agency (ESA) mission {\it Gaia} (\url{https://www.cosmos.esa.int/gaia}), processed by the {\it Gaia} Data Processing and Analysis Consortium (DPAC, \url{https://www.cosmos.esa.int/web/gaia/dpac/consortium}). Funding for the DPAC has been provided by national institutions, in particular the institutions participating in the {\it Gaia} Multilateral Agreement.
\vspace{5mm}
\facility{HST(NICMOS)}

\software{Argus \citep{argus} \url{https://github.com/smfactor/Argus},
          astropy \citep{Astropy2013,Astropy2018},  
          PyMultiNest \citep{Buchner2014},
          MultiNest \citep{Feroz2008,Feroz2009,Feroz2013},
          corner \citep{ForemanMackey2016},
          numpy \citep{numpy},
          SciPy \citep{SciPy}
          }

\bibliography{bib}

\startlongtable
\begin{deluxetable}{hlCCCCCCC}
\tabletypesize{\scriptsize}
    \tablewidth{0pt}
    \tablecaption{Individual $5\sigma$ Contrast Ratio Limits in F110W\label{tab:detLim1}}
    \tablehead{\nocolhead{} & \colhead{} & \multicolumn{7}{c}{Contrast Limit in F110W at Sep. [mas]} \\ \nocolhead{target} & \colhead{Source} & \colhead{55--65} & \colhead{75--85} & \colhead{90--110} & \colhead{160--200} & \colhead{250--300} & \colhead{350--400} & \colhead{450--500}}
\startdata
data/10143/U20004/ &    2M 0004-4044 &    2.3 \pm 0.6, 2.1 \pm 0.5 &         3.35 \pm 0.26 &         4.43 \pm 0.30 &         2.42 \pm 0.24 &           3.7 \pm 0.7 &           4.6 \pm 0.8 &           3.9 \pm 0.5 \\
data/10879/U10018/ &    2M 0024-0158 &                    63 \pm 7 &             115 \pm 5 &             138 \pm 6 &              66 \pm 4 &             126 \pm 5 &            138 \pm 12 &             100 \pm 5 \\
data/10143/U13016/ &    2M 0025+4759 &                             &           1.5 \pm 0.4 &         1.59 \pm 0.13 &                       &         1.52 \pm 0.16 &         2.01 \pm 0.21 &           2.8 \pm 0.5 \\
data/10143/U20029/ &    2M 0036+1821 &                34.5 \pm 3.2 &          65.8 \pm 2.8 &          79.2 \pm 3.1 &          41.3 \pm 1.7 &          75.7 \pm 3.5 &          87.0 \pm 3.3 &              55 \pm 5 \\
data/10143/U20037/ &    2M 0045+1634 &                    55 \pm 6 &              95 \pm 4 &             115 \pm 9 &          62.9 \pm 2.9 &             110 \pm 9 &             126 \pm 5 &              83 \pm 4 \\
data/10143/U20052/ &    2M 0107+0041 &                32.8 \pm 3.4 &              60 \pm 5 &              72 \pm 7 &          37.6 \pm 2.4 &              72 \pm 7 &          72.2 \pm 3.0 &              50 \pm 9 \\
data/10879/U20055/ &    2M 0109+2949 &                   55 \pm 10 &             95 \pm 11 &             115 \pm 8 &             63 \pm 10 &             105 \pm 4 &             121 \pm 9 &              80 \pm 7 \\
data/10143/U20064/ &    2M 0123-4240 &                 6.1 \pm 1.7 &          12.3 \pm 1.4 &          17.9 \pm 2.4 &           9.3 \pm 2.2 &          13.5 \pm 1.9 &          16.3 \pm 1.6 &          13.6 \pm 1.3 \\
\enddata
\tablecomments{Median $5\sigma$ contrast (within the stated separation range) in 4 of 5 calibrators. No entry indicates no significant detection at that separation is possible. Two entries (at tight separations) is caused by a reduction in sensitivity to equal brightness targets and indicates a significant detection is possible \emph{between} the two values. The diffraction limit ($\lambda/D$) is roughly 95 and 147 mas in F110W and F170M, respectively.}
\end{deluxetable}

\startlongtable
\begin{deluxetable}{hlCCCCCCC}
\tabletypesize{\scriptsize}
    \tablewidth{0pt}
    \tablecaption{Individual $5\sigma$ Contrast Ratio Limits in F170M\label{tab:detLim2}}
    \tablehead{\nocolhead{} & \colhead{} & \multicolumn{7}{c}{Contrast Limit in F170M at Sep. [mas]} \\ \nocolhead{target} & \colhead{Source} & \colhead{55--65} & \colhead{75--85} & \colhead{90--110} & \colhead{160--200} & \colhead{250--300} & \colhead{350--400} & \colhead{450--500}}
\startdata
data/10143/U20004/ &    2M 0004-4044 &    6.3 \pm 2.6, 2.01 \pm 0.21 &   16.3 \pm 2.5, 1.32 \pm 0.15 &                  28 \pm 5 &              60 \pm 5 &          26.1 \pm 3.0 &          54.6 \pm 2.1 &                               55 \pm 10 \\
data/10879/U10018/ &    2M 0024-0158 &       33 \pm 7, 1.59 \pm 0.20 &       66 \pm 8, 1.20 \pm 0.10 &                 95 \pm 13 &            201 \pm 34 &            105 \pm 14 &            201 \pm 22 &                              167 \pm 32 \\
data/10143/U13016/ &    2M 0025+4759 &                               &                               &               1.9 \pm 0.5 &           3.7 \pm 0.6 &         1.39 \pm 0.11 &           3.9 \pm 0.4 &                           2.92 \pm 0.14 \\
data/10143/U20029/ &    2M 0036+1821 &       16 \pm 4, 1.83 \pm 0.09 &       31 \pm 4, 1.45 \pm 0.06 &                  60 \pm 9 &              95 \pm 6 &              48 \pm 8 &             105 \pm 4 &                               87 \pm 17 \\
data/10143/U20037/ &    2M 0045+1634 &   20.6 \pm 1.5, 1.83 \pm 0.09 &       45 \pm 7, 1.45 \pm 0.16 &                 79 \pm 14 &             126 \pm 6 &              63 \pm 5 &             138 \pm 5 &                              115 \pm 11 \\
data/10143/U20052/ &    2M 0107+0041 &   17.1 \pm 1.8, 1.75 \pm 0.07 &       38 \pm 4, 1.45 \pm 0.17 &                  66 \pm 7 &             115 \pm 5 &          54.6 \pm 2.3 &             115 \pm 8 &                               100 \pm 5 \\
data/10879/U20055/ &    2M 0109+2949 &       33 \pm 7, 1.59 \pm 0.10 &       66 \pm 8, 1.20 \pm 0.13 &                 138 \pm 9 &            211 \pm 33 &            110 \pm 11 &            231 \pm 11 &  220 \pm 40 \\
\enddata
\tablecomments{Median $5\sigma$ contrast (within the stated separation range) in 4 of 5 calibrators. No entry indicates no significant detection at that separation is possible. Two entries (at tight separations) is caused by a reduction in sensitivity to equal brightness targets and indicates a significant detection is possible \emph{between} the two values. The diffraction limit ($\lambda/D$) is roughly 95 and 147 mas in F110W and F170M, respectively.}
\end{deluxetable}

\startlongtable
\begin{deluxetable}{hlCCChCC}
\tabletypesize{\scriptsize}
\tablewidth{0pt}
\tablecaption{Individual $5\sigma$ Contrast Ratio Limits at Tight Separations\label{tab:detLim3}}
\tablehead{\nocolhead{} & \colhead{} & \multicolumn{3}{c}{F110W Limit at Sep. [mas]} & \multicolumn{3}{c}{F170M Limit at Sep. [mas]}\\ \nocolhead{target} & \colhead{Source} & \colhead{15--25} &\colhead{25--35} & \colhead{35--45} & \nocolhead{15--25} & \colhead{25--35} & \colhead{35--45}}
\startdata
data/10879/U10018/ &    2M 0024-0158 &    4.4 \pm 1.1, 3.1 \pm 0.7 &         8 \pm 4, 2.1 \pm 0.4 &       24 \pm 6, 1.67 \pm 0.08 &                       &      5.9 \pm 1.4, 2.5 \pm 0.6 &   8.9 \pm 2.3, 2.10 \pm 0.30 \\
data/10143/U20029/ &    2M 0036+1821 &                             &   5.3 \pm 2.2, 2.54 \pm 0.27 &       12 \pm 4, 1.92 \pm 0.08 &                       &                               &   4.4 \pm 0.4, 2.92 \pm 0.14 \\
data/10143/U20037/ &    2M 0045+1634 &    4.0 \pm 1.0, 3.7 \pm 0.9 &         9 \pm 4, 2.1 \pm 0.4 &   20.6 \pm 3.5, 1.75 \pm 0.09 &                       &                               &   5.9 \pm 1.0, 2.54 \pm 0.11 \\
data/10143/U20052/ &    2M 0107+0041 &                             &   6.4 \pm 1.0, 2.31 \pm 0.19 &   12.3 \pm 3.2, 2.01 \pm 0.14 &                       &                               &   5.9 \pm 0.9, 2.54 \pm 0.12 \\
data/10879/U20055/ &    2M 0109+2949 &                             &   7.7 \pm 3.2, 2.31 \pm 0.28 &       20 \pm 6, 1.67 \pm 0.08 &                       &      4.9 \pm 1.2, 2.9 \pm 0.4 &   9.9 \pm 3.2, 2.21 \pm 0.18 \\
data/10143/U20064/ &    2M 0123-4240 &                             &                              &      3.7 \pm 0.9, 2.5 \pm 0.6 &                       &                               &                              \\
\enddata
\tablecomments{Median $5\sigma$ contrast (within the stated separation range) in 4 of 5 calibrators. No entry indicates no significant detection at that separation is possible. Two entries (at tight separations) is caused by a reduction in sensitivity to equal brightness targets and indicates a significant detection is possible \emph{between} the two values. The diffraction limit ($\lambda/D$) is roughly 95 and 147 mas in F110W and F170M, respectively.}
\end{deluxetable}

\end{document}